\documentclass[journal]{IEEEtran}%
\ifCLASSINFOpdf
   \usepackage[pdftex]{graphicx}
   \graphicspath{{../pdf/}{../jpeg/}}
   \DeclareGraphicsExtensions{.pdf,.jpeg,.png}
\else
   \usepackage[dvips]{graphicx}
   \graphicspath{{../eps/}}
   \DeclareGraphicsExtensions{.eps}
\fi

\usepackage[nolist,nohyperlinks]{acronym}
\usepackage{amsfonts}
\usepackage{amsmath}
\usepackage{color}
\usepackage{bm}
\usepackage{xcolor}
\usepackage{tabularx}
\usepackage{cite}
\usepackage[ruled,vlined]{algorithm2e}
\usepackage[utf8]{inputenc}
\makeatletter
\newcommand{\nosemic}{\renewcommand{\@endalgocfline}{\relax}}
\newcommand{\dosemic}{\renewcommand{\@endalgocfline}{\algocf@endline}}
\newcommand{\norm}[1]{\left\lVert#1\right\rVert}
\newcommand{\revOne}[1]{\textcolor{black}{#1}}
\newcommand{\revTwo}[1]{\textcolor{black}{#1}}
\newcommand{\arxiv}{}
\newtheorem{definition}{Definition}

\DeclareUnicodeCharacter{2212}{-}

\makeatother

\acrodef{pomdp}[Dec-POMDP]{decentralized partially observable Markov decision process}

\acrodef{posg}[POSG]{partially observable stochastic game}

\acrodef{mdp}[MDP]{Markov decision process}

\acrodef{cmdp}[CMDP]{constrained Markov decision process}

\acrodef{csma}[CSMA]{subcarrier sensing multiple access}

\acrodef{csma_ca}[CSMA/CA]{subcarrier sensing multiple access/collision avoidance}

\acrodef{dql}[DQL]{deep q-learning}

\acrodef{dqn}[DQN]{deep q-network}

\acrodef{dnn}[DNN]{deep neural network}

\acrodef{gfra}[GFRA]{grant-free random access}

\acrodef{ofdma}[OFDMA]{orthogonal frequency division multiple access}

\acrodef{noma}[NOMA]{non-orthogonal multiple access}

\acrodef{ofdm}[OFDM]{orthogonal frequency division multiplexing}

\acrodef{ppo}[PPO]{proximal policy optimization}

\acrodef{trpo}[TRPO]{trust region policy optimization}

\acrodef{rach}[RACH]{random access channel}

\acrodef{dcqn}[DCQN]{deep convolutional Q network}

\acrodef{cnn}[CNN]{convolutional neural network}

\acrodef{iql}[IQL]{independent Q-learning}

\acrodef{ppo}[PPO]{proximal policy optimization}

\acrodef{rl}[RL]{reinforcement learning}

\acrodef{drl}[DRL]{deep reinforcement learning}

\acrodef{marl}[MARL]{multiagent reinforcement learning}

\acrodef{cert}[CERT]{concurrent experience replay trajectories}

\acrodef{bs}[BS]{base station}

\acrodef{tti}[TTI]{transmission time interval}

\acrodef{mac}[MAC]{medium access control}

\acrodef{phy}[PHY]{physical layer}

\acrodef{cw}[CW]{congestion window}

\acrodef{ber}[BER]{bit error rate}

\acrodef{plr}[PLR]{packet loss rate}

\acrodef{qam}[QAM]{quadrature amplitude modulation}

\acrodef{plr}[PLR]{packet loss rate}

\acrodef{awgn}[AWGN]{additive white Gaussian noise}

\acrodef{iid}[i.i.d.]{independent and identically distributed}

\acrodef{der}[DER]{damped exploration renewal}

\acrodef{tfe}[TFE]{trajectory feature extraction}

\acrodef{fifo}[FIFO]{first-in first-out}

\acrodef{sgd}[SGD]{stochastic gradient descent}

\acrodef{mtd}[MTD]{machine-type device}

\acrodef{mmtc}[mMTC]{massive machine-type communication}

\acrodef{phy}[PHY]{physical layer}

\acrodef{h2h}[H2H]{human-to-human}

\acrodef{iot}[IoT]{Internet of Things}

\acrodef{m2m}[M2M]{machine-to-machine}

\acrodef{qos}[QoS]{quality of service}

\acrodef{nb-iot}[NB-IoT]{narrowband internet of things}

\acrodef{amc}[AMC]{adaptive modulation and coding}

\acrodef{wap}[WAP]{wireless access point}

\acrodef{dpm}[DPM]{dynamic power management}

\acrodef{jal}[JAL]{joint action learners}

\acrodef{ann}[ANN]{artificial neural network}

\acrodef{gru}[GRU]{Gated Recurrent Unit}

\acrodef{lstm}[LSTM]{long short term memory}

\acrodef{lte}[LTE]{long term evolution}

\acrodef{urllc}[URLLC]{ultra reliable low-latency communication}

\acrodef{il}[IL]{independent learners}

\acrodef{dacc}[DACC]{distributed actors with centralized critic}

\acrodef{cldi}[CLDI]{centralized learning with decentralized inference}

\acrodef{ack}[ACK]{acknowledgement}

\acrodef{nack}[NACK]{negative-acknowledgement}

\acrodef{adam}[ADAM]{adaptive moment estimation}

\acrodef{lbt}[LBT]{listen before talk}

\acrodef{nr}[NR]{new radio}

\acrodef{harq}[HARQ]{hybrid automatic repeat request}
\begin{document}
%
\title{Intelligent Link Adaptation for Grant-Free Access Cellular Networks: A Distributed Deep Reinforcement Learning Approach}
%
%
%

\author{Joao V. C. Evangelista,~\IEEEmembership{Student Member,~IEEE,}
        ~Zeeshan Sattar,~\IEEEmembership{Member,~IEEE}, Georges Kaddoum,~\IEEEmembership{Senior Member,~IEEE}, Bassant Selim,~\IEEEmembership{Member,~IEEE,} and Aydin Sarraf
\ifdefined \arxiv
\thanks{This work has been submitted to the IEEE for possible publication.  Copyright may be transferred without notice, after which this version may no longer be accessible.

Part of this work was done while the first author was an Mitacs Accelerate intern at Ericsson Canada. This research was also sponsored by a FRQNT B2X Scholarship.

J.V.C. Evangelista, G. Kaddoum and Z. Sattar  were with the Department
of Electrical Engineering, École de Technologie Supérieure, Montreal,
QC, H3C 1K3 CA, e-mail: \{joao-victor.de-carvalho-evangelista.1, zeeshan.sattar.1\}@ens.etsmtl.ca and georges.kaddoum@etsmtl.ca 

B. Selim and A. Sarraf were with the Ericsson Canada, Montreal, QC, H4S 0B6 CA, email: \{bassant.selim, aydin.sarraf\}@ericsson.com 
}
\else
\fi
}
\maketitle

\begin{abstract}
With the continuous growth of machine-type devices (MTDs), it is expected that massive machine-type communication (mMTC) will be the dominant form of traffic in future wireless networks. Applications based on this technology, have fundamentally different traffic characteristics from human-to-human (H2H) communication, which involves a relatively small number of devices transmitting large packets consistently. Conversely, in mMTC applications, a very large number of MTDs transmit small packets sporadically. Therefore, conventional grant-based access schemes commonly adopted for H2H service, are not suitable for mMTC, as they incur in a large overhead associated with the channel request procedure. We propose three grant-free distributed optimization architectures that are able to significantly minimize the average power consumption of the network. The problem of physical layer (PHY) and medium access control (MAC) optimization in grant-free random access transmission is is modeled as a partially observable stochastic game (POSG) aimed at minimizing the average transmit power under a per-device delay constraint. 
The results show that the proposed architectures are able to achieve significantly less average latency than a baseline, while spending less power. Moreover, the proposed architectures are more robust than the baseline, as they present less variance in the performance for different system realizations. 
\end{abstract}

\begin{IEEEkeywords}
grant-free, mMTC, multiple access, reinforcement learning, distributed optimization
\end{IEEEkeywords}
%
\IEEEpeerreviewmaketitle

%
%

\section{Introduction}
\label{sec:intro}

\IEEEPARstart{T}{he} rapid growth of the \ac{iot}, autonomous vehicles, smart grids, and other technologies propelled \ac{m2m} communications into one of the dominant applications in cellular networks \cite{MachinaResearch}. Machine communication have fundamentally different traffic patterns compared to \ac{h2h} ones. In \ac{h2h}, a few users consume and produce large quantities of data, whereas in \ac{m2m} applications a large number of devices generate small amounts of data with diverse \ac{qos} requirements \cite{MachinaResearch}. Given this difference, the grant-based transmission approach adopted by current cellular standards is inefficient in the \ac{mmtc} scenario \cite{Gao2019}. 

A considerable amount of the devices using the \ac{mmtc} service are battery powered, whereas in currently deployed wireless systems, a lot of the energy consumed by communicating devices is used for establishing and maintaining connections. As identified in \cite{Au2014}, when transmitting small packets, the grant request procedure can result in a significant overhead. While the semi-persistent connection, as adopted by the \ac{nb-iot} standard, might reduce the signaling overhead, it can only do so efficiently in the case of periodic traffic arrival \cite{Hoymann2016}. The specifications of 5G \ac{nr} introduced the two-step random access procedure (in the rest of this work referred to as grant-free access) \cite{3gpp.38.213}, which allows the users of the network to transmit their data directly on the \ac{rach} as opposed to the traditional four-step channel request approach used in \ac{lte} (in the rest of this work referred to as grant-based access). This flexibility with regards to the random access mechanism gives us the opportunity to rethink the design of future networks to service \ac{m2m} communications.

A grant-free access mechanism can enable devices to transmit data in an arrive and-go manner in the next available slot. Unlike the current grant-based access mechanism in the LTE uplink, devices using grant-free transmission need not wait for a specific uplink grant from the base station. Such a scheme is more desirable for the two broad \ac{iot} use cases in 5G, namely \ac{mmtc} and ultra-reliable low-latency communications (URLLC), as it enables reduced transmission latency, smaller signaling overhead due to the simplification of the scheduling procedure, and improved energy efficiency (battery life) of the IoT devices with a reduction in signaling and ON time. Grant-free and semi-grant-free transmission are considered for low latency IoT transmission in \cite{TR38.192}. 

Notwithstanding, in grant-free access, as the transmissions are not scheduled on orthogonal time-frequency resources, there is a high probability that different devices will randomly choose the same resource blocks for uplink transmission, resulting in the superposition of data (collision).  The cross-layer optimization of a grant-free network requires consideration of both \ac{phy} and \ac{mac} layer measurements while taking into account their interaction to obtain the relevant performance metrics (e.g. the average power consumption and delay), making it a very challenging problem. Moreover, grant-free transmission poses new challenges in the design of \ac{phy} and \ac{mac} protocols. In this context, static policies for \ac{amc}, power control, and packet retransmission are not able to efficiently satisfy the diverse throughput, latency, and power saving requirements of \ac{mmtc}. Furthermore, due to a lack of scheduling by a central entity in such networks, a distributed optimization approach with partial state information is a natural choice to optimize the network performance while keeping communication overhead to a minimum. In this light, we argue that modeling the problem as a \ac{posg} and proposing a solution within the \ac{marl} framework is the best approach to this problem. While the \ac{posg} model is able to elegantly capture the evolution of the wireless environments and its interactions with the users in time, the \ac{marl} framework enables a distributed decision-making solution, balancing short-term and long-term performance goals.


\subsection{Related Work}
\label{sec:lit_rev}
Random access is an essential component of every multiuser wireless communication system, either as a method of establishing a connection between an user and a \ac{bs} (in grant-based systems), or to transmit data (in grant-free systems). Recently, the rise in prominence of \ac{mmtc} applications has sparked a debate around which of the competing methods should be adopted. A large portion of the research community started advocating for the adoption of a grant-free approach to serve \ac{mmtc} applications \cite{Liu2018, Bockelmann2016}. In \cite{Gao2019} an extensive comparison of the performance of grant-free and grant-based systems with and without subcarrier sensing against a variable packet length is presented. The authors concluded that for shorter packet lengths (as expected in \ac{mmtc} applications) grant-free transmission with sensing results in the best throughput, making a stronger case for grant-free \ac{mmtc} systems. Moreover, in \cite{Evangelista2019, Liu2020}, stochastic geometric models and analytical results for grant-free \ac{noma} are presented.

In \cite{Abreu2018}, the authors propose an open loop power control scheme with path loss compensation in an uplink grant-free \ac{urllc} network to minimize the outage probability. They investigate the effects of the path-loss compensation factor and the expected received power on the network outage probability. In \cite{Jacquelin2019}, the authors introduce a model to abstract multi-packet reception in grant-free networks. They analyze the dynamics of the network and propose a \ac{rl} approach to determine the amount of resource blocks to allocate to grant-free transmission in order to maximize the normalized throughput. In \cite{Huang2019}, the effects of pilot selection in a grant-free \ac{noma} system are investigated, and a \ac{drl} approach is proposed, where each user, without any information exchange, selects its pilots in order to maximize its throughput. Despite their contributions, none of these works address the interactions between the \ac{phy} and \ac{mac} layers and how to harness their flexibility to improve the overall network performance.

In \cite{Mastronarde2013}, a reinforcement learning algorithm is proposed to jointly select an \ac{amc}, and \ac{dpm}, in order to minimize the transmitted power in a single-user system while satisfying a certain delay constraint. Afterwards, in \cite{Mastronarde2016} the work in \cite{Mastronarde2013}  is extended to consider a multiuser system in an IEEE 802.11 network with subcarrier sensing multiple access (CSMA). The authors considered three users contending for channel access, and adopted an independent learners approach \cite{Claus1998}, where each user optimizes its own rewards, ignoring all interactions with other users. Despite its simplicity, the independent learners solution is known to have several issues such as Pareto-selection, nonstationarity, stochasticity, alter-exploration and shadowed equilibria \cite{Matignon2012}.

Although instructive, none of these works addressed the challenges involved in the distributed cross-layer optimization of the grant-free uplink transmission for \ac{mmtc} service. Moreover, previous works on this topic have failed to address the issues involved in the massive scale aspect of \ac{mmtc} applications, despite their contributions. In this manuscript, we propose three distributed solutions based on \ac{marl}, ranging from a fully distributed solution to a centralized learning with distributed inference, to minimize the average power consumption of the network while satisfying delay constraints.


\subsection{Contributions}
\label{sec:contributions}
Although instructive, none of these works addressed the challenges involved in the distributed cross-layer optimization of the grant-free uplink transmission for \ac{mmtc} service. In this manuscript, we propose three distributed solutions based on \ac{marl}, ranging from a fully distributed solution to a centralized learning with distributed inference, to minimize the average power consumption of the network while satisfying delay constraints. The contributions of this paper are summarized as:
\begin{itemize}
    \item We propose a \ac{posg} to model the \ac{phy} and \ac{mac} dynamics of a grant-free \ac{mmtc} network and to formulate the cross-layer power minimization problem. This model considers the channel and packet generation dynamics, and accommodates \acp{mtd} with diverse \ac{qos} requirements and packet arrival intensities.
    \item We propose a fully distribute \ac{il} architecture, based on the \ac{ppo} algorithm \cite{Schulman2015a}, to eliminate the all the communication overhead involved in the cross-layer optimization.
    \item We propose a \ac{dacc} architecture where the \ac{ppo} actor and critic are split and each agent trains its own actor while a single critic is trained by a central entity running on an \revTwo{edge} computing node. This architecture achieves a reduced overhead while allowing the possibility of cooperative behavior to arise among the \acp{mtd}, in our second scheme. As the central entity is able to aggregate measurements from every user, the critic's loss function is calculated from a global performance measure.
    \item We propose \ac{cldi} architecture to eliminate the exponential increase of the policy search space with more \acp{mtd}. In this scheme, every \ac{mtd} uses the same policy, which is trained on an \revTwo{edge} computing node. However, each \ac{mtd} uses the model in a distributed fashion by making decisions based on local data.
    \item We provide an extensive analysis of the performance of all three architectures when servicing \acp{mtd} with diverse \ac{qos} requirements and packet arrival rates. Moreover, we compare their performance with a reactive \ac{harq} protocol with power boosting as a baseline. Finally, we include a quantitative analysis of the tradeoffs involving the performance and the overhead of the proposed architectures in scenarios with different device deployment densities.
\end{itemize}

\subsection{Notation}
\label{sec:notation}
Throughout this paper, italic lowercase letters denote real and complex scalar values. Lower case boldface letters denote vectors, while upper case boldface denote matrices. A lowercase letter with one subscript, $x_i$, represents the $i$-th element of the vector $\mathbf{x}$, while two subscripts $x_{i,j}$ is used to denote the element on the $i$-th row and $j$-th column of matrix $\mathbf{X}$. The operator $E[\cdot]$ denotes the expected value of a random variable. The function $\mathbb{P}(\cdot)$ represents the probability of an event and $\mathbf{x} \sim \mathcal{CN}(\bm{\mu}, \mathbf{K})$, denotes that $\mathbf{x}$ is a circularly symmetric complex Gaussian random vector, with mean $\bm{\mu}$ and covariance matrix $\mathbf{K}$. The notation $x \sim U(\mathcal{X})$ denotes that $x$ is drawn uniformly from the set $\mathcal{X}$. The indicator function takes an event as argument and is equal to one if the event happens and zero otherwise, and is represented by $\mathbf{1}(\cdot)$. Sets $\mathbb{R}$ and $\mathbb{C}$ and are the sets of real and complex numbers, respectively. The set $\mathbb{B} = \{ 0, 1 \}$ represents the binary numbers. A calligraphic uppercase letter, such as $\mathcal{X}$, denotes a set and $|\mathcal{X}|$ is its cardinality. Throughout the paper several variables denote quantities related to a particular user at a given moment in time (e.g. $x_{i,t}$ is related to the $i$-th user on time slot $t$). To avoid cluttering the notation we drop the subscript related to the time and use it only when indexing a variable over multiple periods of time is necessary.

\subsection{Organization}
\revTwo{
This paper is organized as follows: In Section \ref{sec:system_model} we present the system model, discussing in details the dynamics of the environment introducing the optimization problem we aim to solve. In Section \ref{sec:learning_arch}, we present the three distributed learning architectures proposed in this work. In Section \ref{sec:numerical_exps}, the performance of the three proposed architectures is evaluated and the results are discussed. Finally, in Section \ref{sec:conclusions}, we summarize the conclusions.
}
%
%

\section{System Model}
\label{sec:system_model}
In this paper, we consider the problem of designing a distributed link adaptation solution for a grant-free access 5G network providing \ac{mmtc} service. In a grant-free network, there is no guarantee that a transmission attempt is going to be successful. Hence, the usage of a \ac{harq} protocol is essential to guarantee some reliability to the packet transmissions. In the system under analysis, the \acp{mtd} use a reactive \ac{harq} protocol, where after each transmission attempt the device receives either an \ac{ack} feedback, in case the transmission attempt was successfully decoded, or a \ac{nack} feedback, in case the transmission attempt could not be decoded \cite{Mahmood2019}. Notice that we choose the reactive \ac{harq} protocol because the alternative protocols require repeating the same data on every transmission attempt, increasing the transmission power per attempt, which goes against our design objective of minimizing the power expenditure in the network.

We consider a network with $N_U$ \acp{mtd} and $N_B$ base stations randomly located within a circular area of radius $R$. The distance between the $i$-th device and the $j$-th \ac{bs} is denoted by $d_{i,j}$. 
Moreover, each device is associated to its closest \ac{bs}.  We assume there are $N_K$ orthogonal subcarriers reserved for uplink transmission, and $N_P$ orthogonal preambles. In every \ac{tti} the active devices randomly select one out of the $N_P$ available orthogonal preambles, and one subcarrier out of the $N_K$ available to transmit its data on. The orthogonal preambles are used by the network to detect user activity and estimate the \acp{mtd}' channel response. It is worth highlighting that all the variables discussed in this section are associated to a given \ac{tti}; However, for notation convenience, we drop the subscript $t$ used to denote a specific \ac{tti}. Additionally, if $x$ is a variable at time $t$ we use a prime superscript $x^\prime$ to denote the value of the same variable at $t+1$. 

All devices transmit symbols from a \ac{qam} with order $\beta_i \in \{1, \dots, M \}$, where $M$ is the maximum modulation order. Furthermore, before the start of every \ac{tti}, each device has the option to turn off its radio to save power. The radio state is represented by the variable $x_i \in \{0, 1\}$, whenever $x_i = 1$, the radio is on and consumes $P_{ON}$ watts plus whatever power used for the transmission, and when $x_i = 0$, the radio is off and spends $P_{OFF}$ watts. If the radio is on, the device attempts to transmit its data on that particular \ac{tti}, hence, the user must select a transmission power $p_i \in \mathcal{P} = \{\rho_1, \dots, \rho_{\max}\}$. So, the received signal at the $j$-th \ac{bs} on the $k$-th subcarrier is
\begin{equation}
    \label{eq:received_power}
    r_k = \underset{i = 1}{\overset{N_U}{\sum}} x_i \theta_{i,k} \sqrt{p_{i}} h_{i,j,k} d_{i,j}^{-\alpha/2} u_i + w_k,
\end{equation}
where $\theta_{i,k} \in \{ 0, 1 \}$ indicates whether user $i$ is transmitting on subcarrier $k$, $w_{k} \sim \mathcal{CN} (0, N_0)$ is a circularly symmetric complex normal random variable modeling the \ac{awgn} and $\alpha$ is the path loss exponent. The variable $u_i$ is a symbol from a \ac{qam} constellation with order $\beta_i$ and $\norm{u_i}^2 = 1$. Moreover, $h_{i,j,k}$ represents the small-scale fading experienced by the $i$-th user's signal to the $j$-th \ac{bs} on the $k$-th subcarrier. We assume that the channel remains constant during the \ac{tti} duration. To model the relationship between subsequent channel realizations we consider a first-order Gauss-Markov small-scale flat fading model \cite{Evangelista2019a} where
\begin{equation}
\label{eq:fading_model}
   h^{\prime}_{i,j,k} = \kappa h_{i,j,k} + n_{i,j,k},
\end{equation}
where the innovation $n_{i,j,k} \sim \mathcal{CN}(0,1-\kappa^2)$ is a circularly symmetric complex normal random variable. The correlation between successive fading components is given by \cite{Patzold2012}
\begin{equation}
\label{eq:fading_corr}
    \kappa = J_0 \left(2 \pi f_{\max} \Delta_t \right),
\end{equation}
where $f_{\max}$ is the maximum Doppler frequency, $\Delta_t$ is the duration of a single \ac{tti} and $J_0$ is the zero-th order Bessel function of the first kind.

In order to guarantee a harmonious access to the channel and avoid congestion, the system under investigation employs a rate-adaptive \ac{lbt} mechanism with random backoff on every subcarrier to control the congestion. This approach is well aligned with the specifications of 5G networks operating on the unlicensed spectrum \cite{Dahlman2018book, Kim2020, Maldonado2020, Song2019}. A \ac{tti} is divided into two phases: contention and transmission. During the contention phase, the device listens to the channel on a specific subcarrier for a random backoff time $\tau^C < \Delta_t$. If no other user has started transmission during this time, the device starts its transmission for an amount of time $\tau^{TX} = \Delta_t - \tau^C$. The protocol is illustrated in Fig. \ref{fig:csma_backoff}, where we show a situation in which $4$ devices are transmitting on the same subcarrier. The red shaded areas indicate the random backoff time $\tau^C$ drawn by each user. In this figure, as device 2 drew the smallest backoff time, it takes hold of the channel and transmits its data in the remaining time available in the time slot. In this model, a collision occurs if two devices draw the same random backoff time.

\ifCLASSOPTIONtwocolumn
\begin{figure}
    \centering {
    \includegraphics[width=0.6\columnwidth]{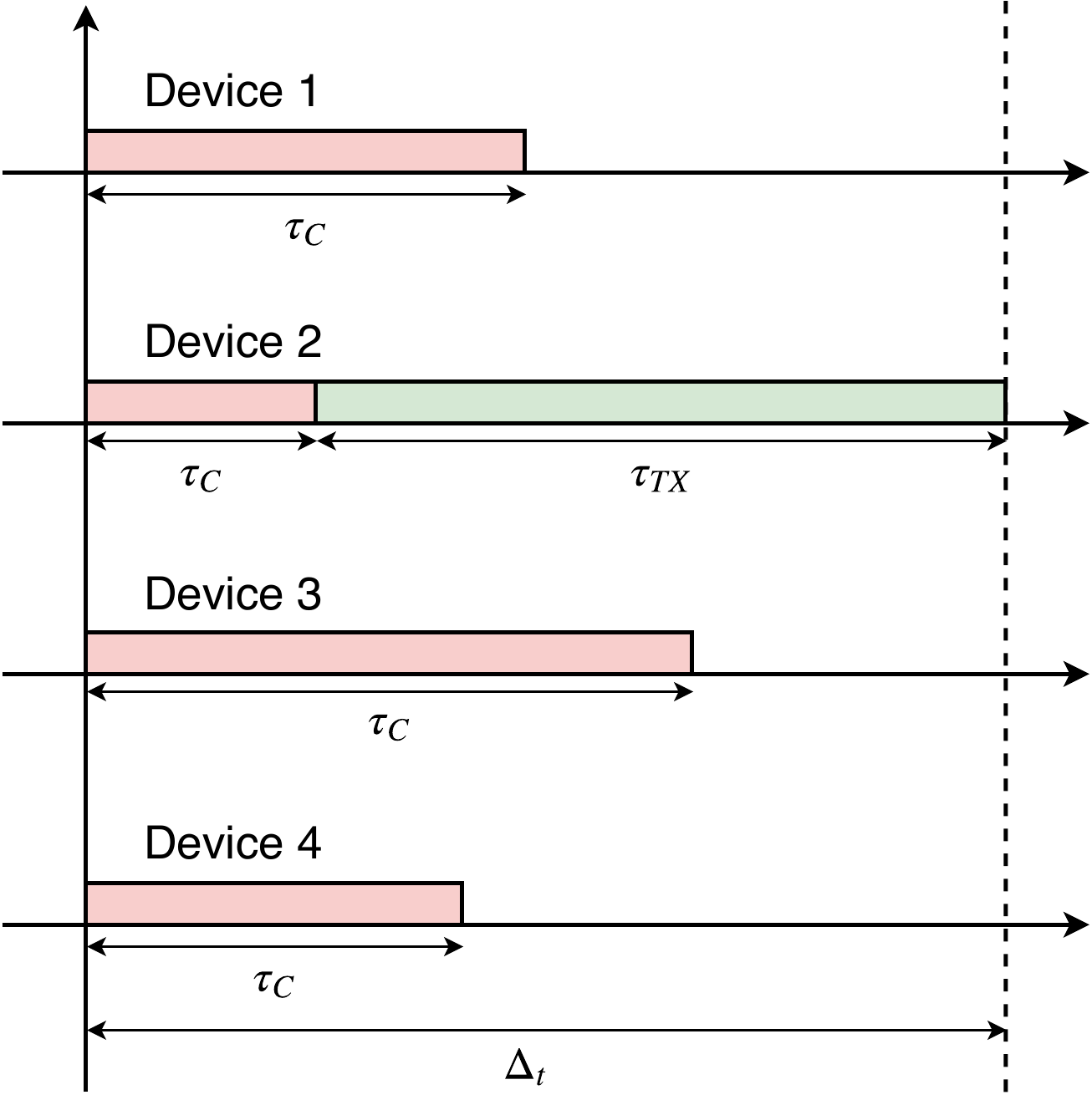}
    \caption{Illustration of the considered \ac{lbt} procedure for $4$ devices sharing the same channel. The red shaded area represents the random backoff listening time and the green shaded one denotes the transmission time.}
    \label{fig:csma_backoff}
    }
\end{figure}
\else
\begin{figure}
    \centering {
    \includegraphics[width=0.38\columnwidth]{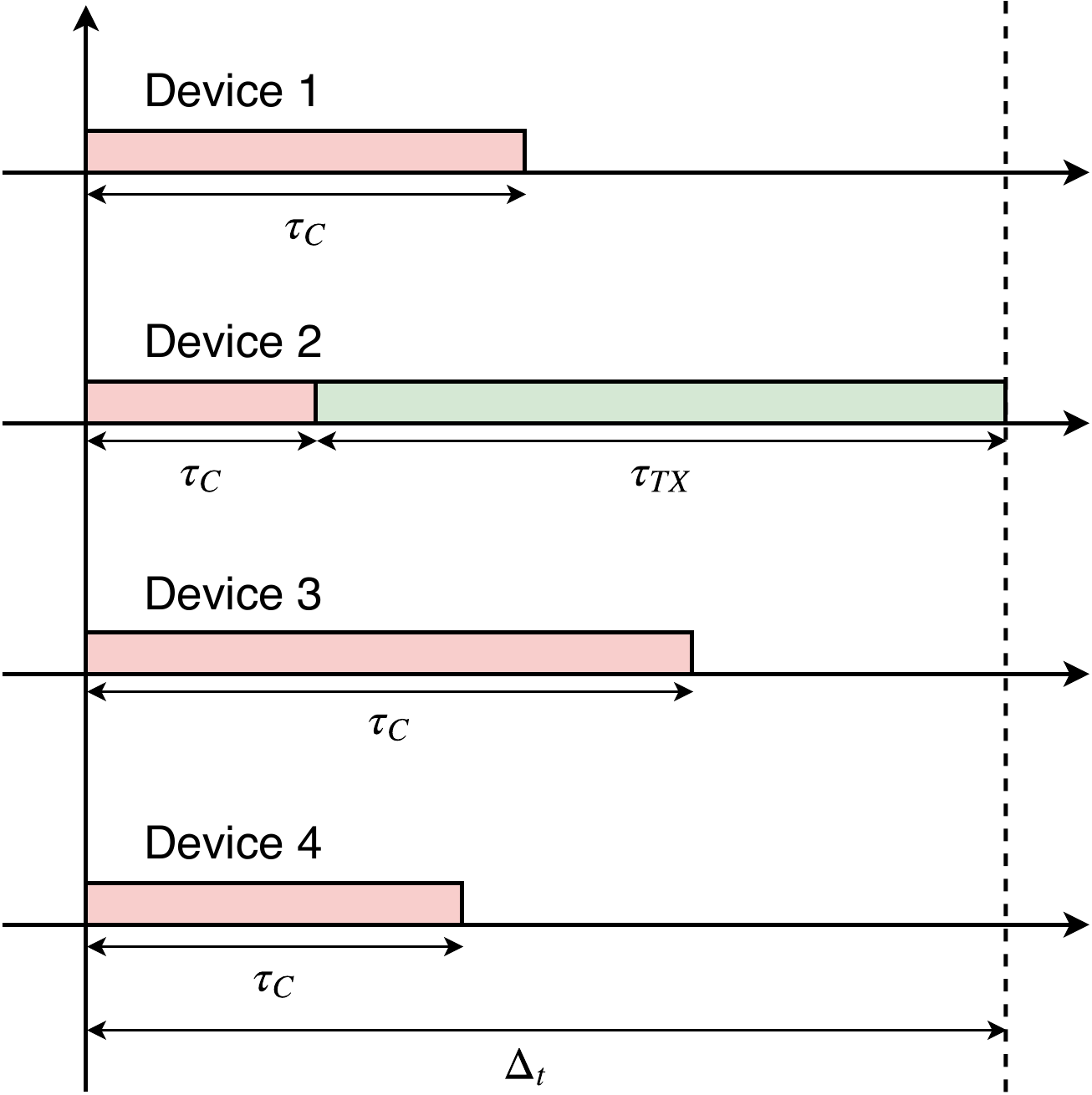}
    \vspace{-.7cm}
    \caption{Illustration of the considered \ac{lbt} procedure for $4$ devices sharing the same channel. The red shaded area represents the random backoff listening time and the green shaded one denotes the transmission time.}
    \label{fig:csma_backoff}
    }
\end{figure}
\fi

We consider a rate-adaptive congestion control protocol, similar to the one proposed in \cite{Mastronarde2016}, where a \ac{cw}, given by $CW_{\min}(\beta_i) = \lfloor A 2^{M - \beta_i} \rfloor$, where $A \in \mathbb{R}$ is a design parameter, is assigned to the device according to its modulation order. The backoff time of the $i$-th device is uniformly chosen from $[0, CW_{\min}(\beta_i)]$ and is reset at the end of the time slot. 
\begin{definition}[Collision]
    We consider that a collision occurs whenever two devices being served by the same \ac{bs} select the same preamble and the same subcarrier, and, draw the same random backoff time $\tau^C$.
\end{definition}
If a collision occurs, the devices' \ac{cw} are set to $CW_{\max} = A 2^M$. Note that the \ac{mtd} attempts to transmit
\begin{equation}
    \label{eq:num_packets}
    z_i = \left \lfloor \frac{\beta_i \tau^{TX}}{L T_S} \right \rfloor
\end{equation}
packets in a given \ac{tti}, where $L$ is the packet length and $T_S$ is the symbol duration. This approach increases the likelihood that a device that intends to transmit at higher rates obtains channel access, avoiding the anomaly identified in \cite{Heusse2003a}, where low-rate users significantly degrade the performance of the whole network. 

Furthermore, we assume that each device has a packet buffer with a capacity of $L_B$ packets. Let $b_i$ be the number of packets in the $i$-th device's buffer. We assume that the number of arriving packets follow a Poisson distribution $l_i \sim Poisson(\lambda_i)$, where $\lambda_i$ is th mean packet arrival rate. The number of packets departing (the goodput) the device's buffer is denoted by $g_i$. The goodput of the $i$-th \ac{mtd} is a function of the device's transmit power, its selected subcarrier, its channel to the receiving \ac{bs}, and the interference power at the receiving \ac{bs}. Let the interference suffered by the $i$-th \ac{mtd}'s transmission on the $k$-th subcarrier be
\begin{equation}
    I_{i,k} = \underset{\underset{n \neq i}{n = 1}}{\overset{N_U}{\sum}} x_n \theta_{n,k} p_{n} \norm{h_{n,j,k}}^2 d_{n,j}^{-\alpha}
\end{equation}
The probability that the $j$-th \ac{bs} decodes a bit transmitted by the $i$-th \ac{mtd} in error (denoted as $P_i^{e}$) can be approximated by \cite{Proakis2007}
\begin{eqnarray}
\label{eq:ber}
    P_i^{e}
    \approx \begin{cases}
        \frac{1}{2} \text{erfc} \left\{ \sqrt{\frac{p_i \norm{h_{i, j, k}}^2 d_{i, j}^{-\alpha}}{I_{i,k} + N_0}} \right\} & \text{if } \beta_i = 1 \\
        2 \text{erfc} \left\{ \sqrt{\frac{3 \log_2(\beta_i) p_i  \norm{h_{i, j, k}}^2 d_{i, j}^{-\alpha} }{2 (\beta_i-1) (I_{i,k} + N_0)} }\right\} & \text{if } \beta_i > 1,
    \end{cases}
\end{eqnarray}
Given the approximate probability of decoding a bit in error given in \eqref{eq:ber}, we obtain the probability of losing a packet as
\begin{equation}
\label{eq:PLR_from_BER}
    P^{loss}_i = 1 - \left( 1-P_i^e \right)^L.
\end{equation}
Moreover, the number of overflown packets, i.e. packets that arrive while the buffer is full and must be dropped, at the $i$-th device's buffer is given by
\begin{equation}
\label{eq:overflow_packets}
   \xi_i = \max(b_i + l_i - g_i - L_B, 0).
\end{equation}

\subsection{Problem Formulation}
\label{sec:problem_formulation}
The main goal of this work is to derive a link adaptation algorithm to minimize the average power consumption over time of the network under a constraint on the average delay. Notice that as the \acp{mtd} transmission attempts are not scheduled by a central network the proposed algorithm must run on each device in a distributed fashion. Also, although the goal is to minimize the average power consumption the algorithm has only local information to make decisions on the link adaptation. This problem can be formulated as a \ac{posg} \cite{Neyman2003}. A \ac{posg}, models how multiple agents, with distinct and possibly adversarial goals, interact with a stochastic changing environment in discrete time slots. At each time slot, the agents receive a partial, and possibly noisy, observation of the environment and select an action to take in the next slot based on this observation. Each set of actions selected by the agents incurs a cost and the objective of the problem is to find the joint policy that minimizes the cost. In this work, we are concerned with infinite horizon \ac{posg}s \cite{Puterman2014}, as the task we are optimizing cannot be described by finite length episodes. The \ac{posg} problem is formally defined by a tuple $\left(\mathcal{U}, \mathcal{S}, \mathcal{A}, \mathcal{P}_{\mathcal{S}}, c, \mathcal{O} \right)$, where $\mathcal{U}$ is the set of agents, where each \ac{mtd} out of the $N_U$ total constitutes an agent. $\mathcal{S}$ and $\mathcal{A} = \times_{i \in \mathcal{U}} \mathcal{A}_i$ denote the state space and the joint action space of the system, respectively, where $\mathcal{A}_i$ is the action space of the $i$-th agent. The state-action transition probability $\mathcal{P}_{\mathcal{S}}:\mathcal{S}\times \mathcal{A} \times \mathcal{S} \rightarrow [0,1]$ gives the probability $\mathcal{P}(\mathbf{s}^\prime|\mathbf{s},\mathbf{a})$ of transitioning to a state $\mathbf{s}^\prime$, given the current state $\mathbf{s}$ and the joint selected action $\mathbf{a}$. The set $\mathcal{O} = \{\mathcal{O}_i : \mathcal{O}_i \subseteq \mathcal{S} \text{ } \forall i \in \mathcal{U}\}$ contains the observation space of each device, which is a subset of the complete state space. Furthermore, $c:\mathcal{S} \times \mathcal{A} \rightarrow \mathbb{R} \text{ }$ is the cost function associated to the problem. The cost function gives the cost of taking action $a$ while on state $s$.
\begin{definition}[Policy]
A policy $\pi(a \lvert o)$, for $a \in \mathcal{A}$ and $o \in \mathcal{O}_i$ is a conditional probability distribution that gives the probability that the agent selects the action $a$ given that it observes the local observation $o$.
\end{definition}
The joint policy of all the agents is denoted by $\bm{\pi} = [ \pi_1, \dots, \pi_{N_U} ]$. Notice that $\bm{\pi}$ is also a conditional probability function given by
\begin{equation}
    \pi(\mathbf{a}\lvert \mathbf{s}) = \underset{i = 1}{\overset{N_U}{\prod}} \pi_i(a_i \lvert o_i)
\end{equation}
The optimality criteria defines the optimization objective of the problem. In the case of an infinite-horizon \ac{posg}, we want the average cost over time to be minimized. Therefore, a natural optimality criteria for the joint policy $\bm{\pi}$ is the expected discounted cost \cite{Oliehoek2016}, which is given by
\begin{equation}
    \label{eq:discounted_cost}
    C_\pi(\mathbf{s}) = E_{\bm{\pi}} \left[\left.  \underset{t = 0}{\overset{\infty}{\sum}} \gamma^t c(\mathbf{s}_t, \mathbf{a}_t) \right\rvert \mathbf{s}_0 = \mathbf{s} \right],
\end{equation}
where $0 < \gamma < 1$ is the discount factor. So, the cost function takes into account the effect of the action on the current and future \acp{tti}. The discount factor is necessary to keep the summation in (\ref{eq:discounted_cost}) bounded and can be interpreted as how much weight should the agent's decision give to future costs. 

Let $\bm{\Pi} = \{\bm{\pi} \lvert \bm{\pi}:\mathcal{A} \times \mathcal{S} \rightarrow [0, 1] \}$ be the set of all possible joint policies. Then, the solution of a \ac{posg} is defined as
\ifCLASSOPTIONtwocolumn
\begin{eqnarray}
    \label{eq:posg_solution}
    \bm{\pi}^* &=&  \arg \underset{\bm{\pi} \in \bm{\Pi}}{\min} \text{ } E_{\mathbf{s} \sim \mathbb{P}(\mathbf{s})} \left[\left. C_{\pi}(\mathbf{s}) \right\rvert \mathbf{s}_0 = \mathbf{s} \right] \nonumber \\ 
    &=& \arg \underset{\bm{\pi} \in \bm{\Pi}}{\min} \text{ }  E_{\mathbf{s} \sim \mathbb{P}(\mathbf{s}), \bm{\pi}} \left[\left.  \underset{t = 0}{\overset{\infty}{\sum}} \gamma^t c(\mathbf{s}_t, \mathbf{a}_t)\right\rvert \mathbf{s}_0 = \mathbf{s} \right], \nonumber \\ &&
\end{eqnarray}
\else
\begin{equation}
    \label{eq:posg_solution}
    \bm{\pi}^* =  \arg \underset{\bm{\pi} \in \bm{\Pi}}{\min} \text{ } E_{\mathbf{s} \sim \mathbb{P}(\mathbf{s})} \left[\left. C_{\pi}(\mathbf{s}) \right\rvert \mathbf{s}_0 = \mathbf{s} \right] = \arg \underset{\bm{\pi} \in \bm{\Pi}}{\min} \text{ }  E_{\mathbf{s} \sim \mathbb{P}(\mathbf{s}), \bm{\pi}} \left[\left.  \underset{t = 0}{\overset{\infty}{\sum}} \gamma^t c(\mathbf{s}_t, \mathbf{a}_t)\right\rvert \mathbf{s}_0 = \mathbf{s} \right],
\end{equation}
\fi
where $\mathbb{P}(\mathbf{s})$ is the probability distribution over the set of states $\mathcal{S}$ while following the joint policy $\bm{\pi}$, and, $\bm{\pi}^*$ is the policy that minimizes the expected discounted cost from the set of all possible policies. The problem in (\ref{eq:posg_solution}) is known to be undecidable, meaning that given a threshold, it is not possible to tell whether there exists a policy that has an expected discounted cost smaller than the threshold \cite{Madani1999}; However, as we show in Section \ref{sec:learning_arch}, we can reformulate the problem in (\ref{eq:posg_solution}) to a proxy problem, and approximate the policies $\pi_i$ by a parametric function approximator $\pi_{\mathbf{w}_i}$, where $\mathbf{w}_i$ is the set of parameters for the device's policy. Consequently, the set of all possible joint policies $\bm{\Pi}$ becomes constrained to the set of all possible policies that can be approximated by the parametric model. Considering a differentiable parametric model, we can use a data-driven learning approach to optimize the parameters and obtain high-quality sub-optimal solutions to (\ref{eq:posg_solution}).

The cellular system model described so far can be conveniently mapped into the \ac{posg} problem formulation. The state of the system can be denoted by
\begin{equation}
    \label{eq:state}
    \mathbf{s} = \left( \mathbf{h}, \mathbf{b}, \mathbf{l}, \mathbf{g} \right),
\end{equation}
where $\mathbf{h} = \text{vec} \left(\left[ \mathbf{H_1}, \dots, \mathbf{H_{N_U}} \right]\right)$ and $\mathbf{H}_i = [\norm{h_{i, j, k}}^2 d_{i,j}^{-\alpha}]_{j, k}$ is a $(N_B \times N_S)$ matrix where each entry is the channel gain between the $i$-th \ac{mtd} and the $j$-th \ac{bs} on the $k$-th subcarrier. Additionally, $\mathbf{b}$, $\mathbf{l}$, and $\mathbf{g}$ are vectors containing the number of packets in the buffer, the number of arriving packets and the goodput of each \ac{mtd}, respectively. As the devices only have access to their local information the observation vector is given as
\begin{equation}
    \label{eq:observation}
    \mathbf{o}_i = \left[ \text{vec} \left( \mathbf{H}_i \right), x_i, b_i, l_i, g_i \right].
\end{equation}
Furthermore, we map the optimization variables of the power minimization problems into the joint action vector as
\begin{equation}
    \label{eq:action}
    \mathbf{a} = (\bm{\theta}, \bm{\beta}, \mathbf{p}, \mathbf{x}),
\end{equation}
where $\bm{\theta} = \left[ \bm{\theta}_1, \dots, \bm{\theta}_{N_U} \right]$ and $\bm{\theta}_i \in \{0, 1\}^{N_S}$ is the subcarrier selection vector of the $i$-th user, and, $\underset{N_S}{\overset{k = 1}{\sum}} \theta_{i, k} \leq 1$. Also, $\bm{\beta}$, $\mathbf{p}$, $\mathbf{x}$ correspond to the modulation order, power and radio state selected by each \ac{mtd}, respectively.

In this work, we want to minimize the power usage subject to a latency constraint. The \ac{posg} problem formulation is not compatible with a constrained objective. Hence, we follow the approach in \cite{Altman1999} to model contrained Markov decision processes \acused{cmdp}(CMDPs) and augment the objective function with a Lagrangian penalty \cite{Nocedal}. Furthermore, according to Little's theorem \cite{El-Taha2012}, the average number of packets queued in the buffer is proportional to the average packet delay in queues with stable buffers (i.e. no overflow). Hence, we design the cost function to discourage large number of packets in the queue, which we refer to as the holding cost, while simultaneously penalizing dropped packets, which we refer to as the overflow cost. Therefore, in the \ac{posg} formulation, the cost function is
\ifCLASSOPTIONtwocolumn
\begin{eqnarray}
    c(\mathbf{s}, \mathbf{a}) = &\underset{i=1}{\overset{N_U}{\sum}}& \underbrace{x_i (P_{ON} + p_i) + (1 - x_i) P_{OFF}}_{\text{power cost}} + \nonumber \\ &&\omega_i \left( \underbrace{b_i}_{\text{holding cost}} +  \underbrace{\mu \xi_i}_{\text{overflow cost}} \right),
\end{eqnarray}
\else
\begin{equation}
    c(\mathbf{s}, \mathbf{a}) = \underset{i=1}{\overset{N_U}{\sum}} \underbrace{x_i (P_{ON} + p_i) + (1 - x_i) P_{OFF}}_{\text{power cost}} + \omega_i \left( \underbrace{b_i}_{\text{holding cost}} +  \underbrace{\mu \xi_i}_{\text{overflow cost}} \right),
\end{equation}
\fi
where $\omega_i$ is a Lagrange multiplier. Thus, if the $i$-th \ac{mtd} has a delay constraint equal to $\delta_i$, then, $\omega_i \propto \max (0, [b_i + \mu \xi_i] - \delta_i)$ is proportional to how much the delay constraint is being violated. Moreover, $\mu$ is the overflow penalty factor. The overflow penalty factor must be chosen such that dropping packets is sub-optimal, while encouraging devices to transmit with low-power. To meet these requirements, we choose a value of $\mu$ such that dropping a packet costs as much as the largest possible discounted expected cost incurred by holding a packet in the buffer, which happens if the packet is held in the buffer forever. Therefore
\begin{equation}
\label{eq:mu_def}
    \mu = \underset{t = 0}{\overset{\infty}{\sum}} \gamma^{t+1} = \frac{\gamma}{1-\gamma}.
\end{equation}
\section{Distributed Learning Architectures}
\label{sec:learning_arch}
Finding the optimal police to the proposed infinite-horizon \ac{posg} problem is undecidable. Deep neural networks \acused{dnn}(DNNs) are universal function approximators and can be trained to learn a mapping from data efficiently through gradient descent and backpropagation \cite{Goodfellow2016}. Thus, we can use \acp{dnn} to approximate the policies and use the agents' experience to learn policies that minimize the cost. This deep \ac{marl} has been proven to be successful in many complex multiagent tasks \cite{Foerster2016a, Foerster2017b, Foerster2018, Lowe2017, Omidshafiei2017a}. However, many of the problems traditionally investigated in the \ac{marl} literature can be trained on computer clusters, where the computing nodes are connected together through high-speed network interconnections and can easily share information among themselves to mitigate the partial observability of \acp{posg} \cite{Claus1998}. On the other hand, when the computing nodes (in our case \acp{mtd} and edge computing infrastructure) are connected via wireless links, sending additional information incurs in an expensive overhead. Therefore, it is imperative to propose novel ways to train these \acp{dnn} to solve the \ac{posg} problem, while sharing as little information between the computing nodes as possible. For this reason, we chose an actor-critic policy gradient approach \cite{Sutton2018}, as we have more flexibility on distributing the training and inference by placing the actor and the critic on different computing nodes. 

In this setting, we propose three different distributed learning architectures:
\ac{il}, \ac{dacc} and \ac{cldi}. Fig. \ref{fig:arch_comp} illustrates the main differences between these architectures. Firstly, in the \ac{il} architecture, each \ac{mtd} has its own network for policy selection (the actor) and value estimation (the critic). Secondly, in the \ac{dacc}, the value estimator and policy selection networks are decoupled. Each \ac{mtd} has its own policy selection network and an \revTwo{edge} agent, which we assume is connected to every \ac{bs} and has access to the state of every \ac{mtd}, stores and trains a value estimator network. At each \ac{tti} the \revTwo{edge} node feedbacks the critic value of the current state to all \acp{mtd} through a broadcast channel. \acp{mtd} use the fedback value estimate as the actor-critic's baseline to train their policy selection network. In the \ac{cldi} architecture, we follow a similar approach  to \cite{Lowe2017, Foerster2018}, and consider that the \revTwo{edge} node trains the weights (only from local observations) of a single policy network that is shared among all agents and sends it periodically through a broadcast channel. Then, \acp{mtd} are able to select their actions only from local observations. Notice that in the \ac{dacc} and \ac{cldi} architectures, the \acp{mtd} need to feedback their state information back to the \ac{bs}. This can be achieved by appending the buffer information to the transmitted packets, or by scheduling periodic state information transmission through a collision free channel. Nevertheless, in this paper, our aim is to evaluate the performance of the proposed architectures, and thus, we assume the state information can be reliably transmitted to the \ac{bs}. Each approach presents its own advantages and challenges, as detailed in the rest of this section. 
\ifCLASSOPTIONtwocolumn
\begin{figure*}
    \centering {
    \includegraphics[width=1.5\columnwidth]{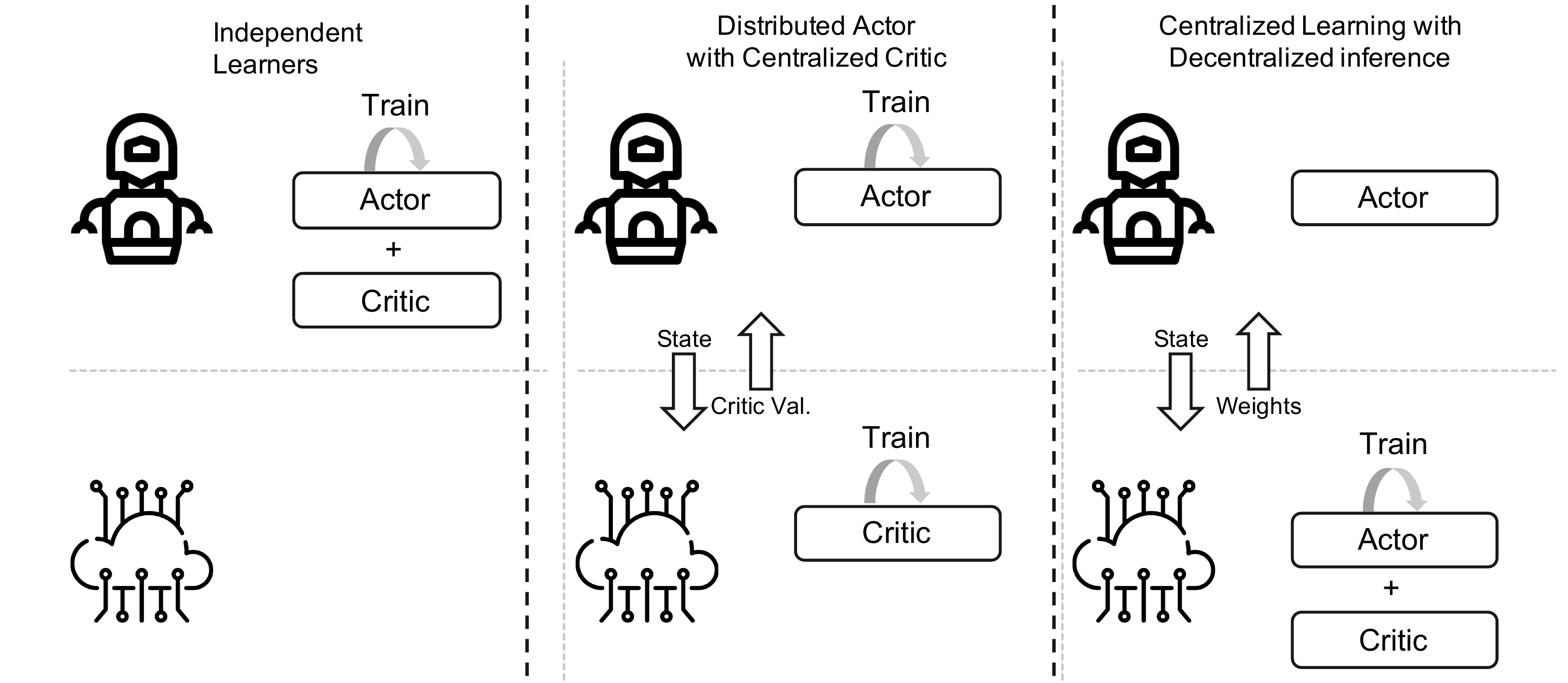}
    \vspace{-.2cm}
    \caption{The diagram illustrates the differences between the three proposed architectures.}
    \label{fig:arch_comp}
    }
\end{figure*}
\else
\begin{figure}
    \centering {
    \includegraphics[width=.75\columnwidth]{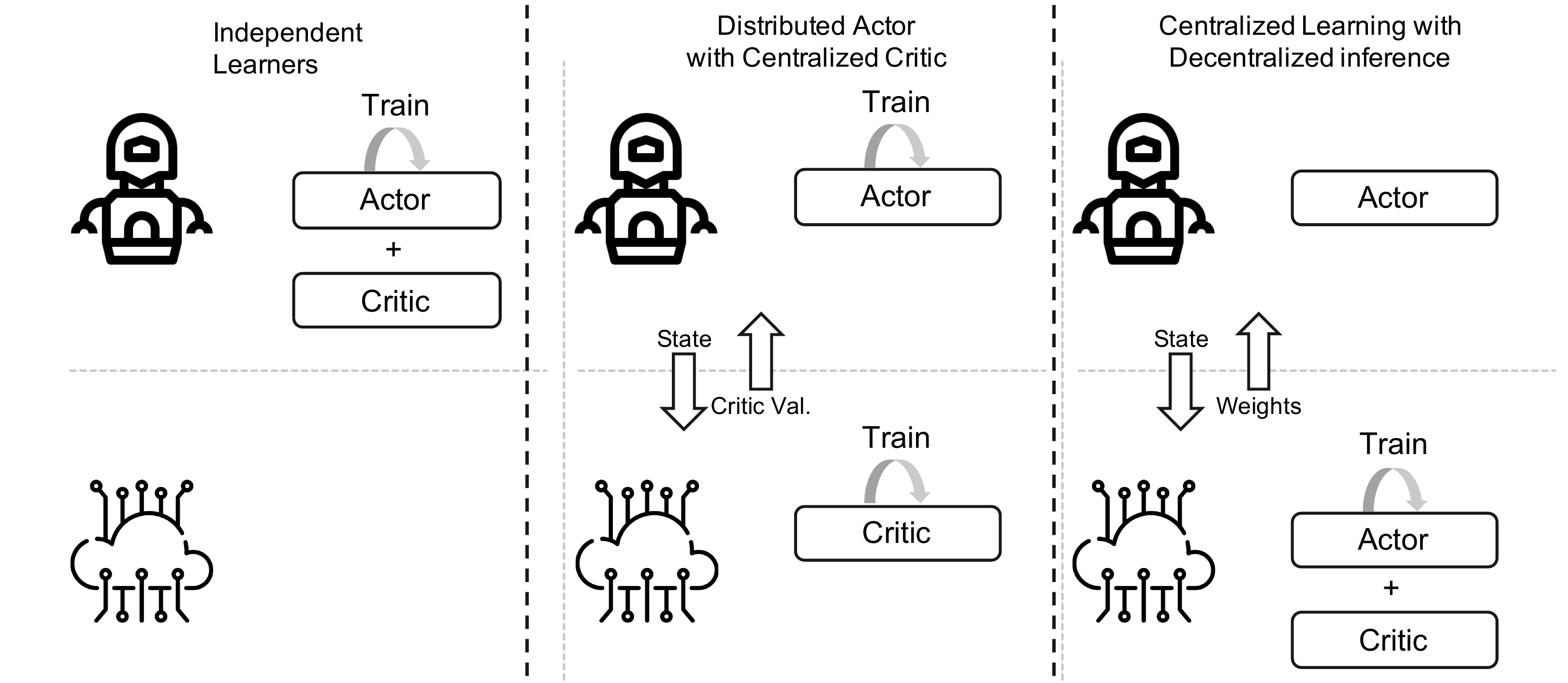}
    \vspace{-.5cm}
    \caption{The diagram illustrates the differences between the three proposed architectures.}
    \label{fig:arch_comp}
    }
\end{figure}
\fi

In order to provide a fair comparison, in all of the proposed architectures, we consider an actor-critic style \ac{ppo} algorithm \cite{Schulman2017}, due to its ease of implementation, the possibility of decoupling the policy and the value estimator, reduced sample complexity compared to \ac{trpo} \cite{Schulman2015a}, and first-order updates. We start this section by introducing policy gradient methods and the \ac{ppo} algorithm, and proceed to describe the three proposed architectures in detail.

\subsection{Policy Gradient Methods}
\label{sec:pg_methods}
In contrast to action-value methods, such as Q-learning \cite{Watkins1992a}, where the agent learns an action-value function and derives a policy from selecting the actions that maximize its output, policy gradient methods learn a parametrized policy that selects the actions without consulting a value function. Let $\mathbf{w} \in \mathbb{R}^d$ be the policy parameter vector, then the parametrized policy $\pi_{\mathbf{w}}(a|s) = \mathbb{P}(a|s,\mathbf{w})$ denotes the probability of selecting action $a$, while at state $s$ with policy parameter $\mathbf{w}$. 

In order to learn the policy parameter vector, we need to have an objective function of $\mathbf{w}$ to be maximized. Consider a scalar performance function $J(\mathbf{w})$, differentiable with respect to $\mathbf{w}$. Then, the learning procedure consists in maximizing $J(\mathbf{w})$ through gradient ascent updates of the form \cite{Sutton2018}
\begin{equation}
\label{eq:weight_update}
    \mathbf{w}^\prime = \mathbf{w} + \eta \nabla_{\mathbf{w}} \tilde{J}(\mathbf{w}),
\end{equation}
where $\eta$ is the learning rate, and $\nabla_{\mathbf{w}} \tilde{J}(\mathbf{w})$ is an estimator of the gradient of the performance measure. A common choice of performance measure is
\begin{equation}
\label{eq:perf_measure}
    J(\mathbf{w}) = \pi_{\mathbf{w}}(a|s) A_{\pi}(s,a),
\end{equation}
where $A_{\pi}(s,a) = Q_{\pi}(s,a) - V_{\pi}(s)$ is the advantage function, which gives the advantage of taking action $a$ while at state $s$ in comparison to the state value function, which gives the value of the average action. The state value function for policy $\pi$, $V_{\pi}(s)$, is given by the expected discounted reward of state $s$ while following policy $\pi$, defined as
\begin{equation}
    \label{eq:value_func}
    V_{\pi}(s) = E_\pi \left[\left. \underset{t = 0}{\overset{\infty}{\sum}} - \gamma^t c(s_t, a_t) \right\lvert s_0 = s\right].
\end{equation}
Furthermore, the action-value function for policy $\pi$, $Q_{\pi}(s, a)$, gives the expected discounted reward of taking action $a$ while in state $s$ and then continuing to follow policy $\pi$, which is given as
\begin{equation}
    \label{eq:action_value_func}
    Q_{\pi}(a, s) = E_\pi \left[\left. \underset{t = 0}{\overset{\infty}{\sum}} - \gamma^t c(s_t, a_t) \right\lvert s_0 = s, a_0 = a \right].
\end{equation}
Notice that both (\ref{eq:value_func}) and (\ref{eq:action_value_func}) can be estimated from experience. 
This class of algorithms are known as \textit{actor-critic} because we evaluate the difference between the actor estimate ($Q_{\pi}(s,a)$) and the critic estimate ($V_{\pi}(s)$), as presented in (\ref{eq:perf_measure}). 

Let $\mathcal{E}_{\pi_{\mathbf{w}}}$ be the set of the experience tuples collected while following policy $\pi_{\mathbf{w}}$, where an experience tuple consists of the state, action, and cost. Then, the gradient of the performance measure can be estimated by taking the average gradient over a random finite batch of experience tuples as
\begin{equation}
\label{eq:perf_gradient}
    \nabla_{\mathbf{w}} \tilde{J}(\mathbf{w}) = \hat{E}_{\mathcal{E}_{\pi_{\mathbf{w}}}} \left[ \nabla_{\mathbf{w}} \ln \pi_{\mathbf{w}}(a|s) A(s_t,a_t) \right],
\end{equation}
where $\hat{E}_{\mathcal{E}_{\pi_{\mathbf{w}}}}$ denotes the empirical average over a batch of randomly sampled experience tuples. 

\subsection{Proximal Policy Optimization}
\label{sec:ppo}
The \ac{ppo} algorithm, originally proposed in \cite{Schulman2017}, consists in maximizing a clipped surrogate objective $J^{\text{clip}}(\mathbf{w})$ instead of the original performance measure $J(\mathbf{w})$, therefore avoiding the destructively large updates experienced in policy gradient methods without clipping as shown in \cite{Schulman2017}. The surrogate objective is defined as 
\ifCLASSOPTIONtwocolumn
in (\ref{eq:ppo_surrogate}), shown on the top of the next page,
\begin{figure*}
\begin{equation}
\label{eq:ppo_surrogate}
    J^{\text{clip}}(\mathbf{w}) = \hat{E}_{\mathcal{E}_{\pi_{\mathbf{w}}}} [ \min ( \Gamma(\mathbf{w}) A(s,a),
    \text{clip}(\Gamma(\mathbf{w}), 1-\epsilon, 1+\epsilon)A(s,a))]
\end{equation}
\hrulefill
\end{figure*}
\else
\begin{equation}
\label{eq:ppo_surrogate}
    J^{\text{clip}}(\mathbf{w}) = \hat{E}_{\mathcal{E}_{\pi_{\mathbf{w}}}} [ \min ( \Gamma(\mathbf{w}) A(s,a),
    \text{clip}(\Gamma(\mathbf{w}), 1-\epsilon, 1+\epsilon)A(s,a))],
\end{equation}
\fi
where $\Gamma(\mathbf{w}) = \frac{\pi_{\mathbf{w}}(a|s)}{\pi_{\mathbf{w}_{\text{old}}}(a|s)}$ is the importance weight, $\epsilon$ is a hyperparameter that controls the clipping, and $\mathbf{w}_{\text{old}}$ are the policy weights prior to the update. Due to the term $\text{clip}(\Gamma(\mathbf{w}) A(s,a), 1-\epsilon, 1+\epsilon)$ in (\ref{eq:ppo_surrogate}), the importance weight is clipped between $1-\epsilon$ and $1+\epsilon$, minimizing the incentives for large destabilizing updates. Furthermore, by taking the minimum of the clipped and unclipped functions, the resulting surrogate objective is a lower bound first-order approximation of the unclipped objective around $\mathbf{w}_{\text{old}}$.

Furthermore, the performance measure is augmented to include a value function loss term, corresponding to the critic output, given by
\begin{equation}
    J^{\text{VF}}(\mathbf{w}) = \hat{E}_{\mathcal{E}_{\pi_{\mathbf{w}}}} \left[ \left( V_{\pi_{\mathbf{w}}}(s) - \underset{k=0}{\overset{|\mathcal{E}|-1}{\sum}} \gamma^k r \right)^2 \right]. 
\end{equation}

Finally, a final term of entropy bonus $H(\pi_{\mathbf{w}})$ is added to encourage exploration of the state space \cite{Mnih2016AsynchronousLearning}. The final surrogate objective function to be maximized is given by
\begin{equation}
\label{eq:ppo_surrogate_final}
    J^{\text{surr}}(\mathbf{w}) = J^{\text{clip}}(\mathbf{w}) - k_1 J^{\text{VF}}(\mathbf{w}) + k_2 H(\pi_{\mathbf{w}}),
\end{equation}
\ifCLASSOPTIONtwocolumn
\begin{figure*}[ht!]
\begin{equation}
    \label{eq:il_posg}
    \pi^{*}_i = \arg \underset{\pi_i \in \Pi^{IL}_i}{\min} \text{ } E_{\mathbf{o}_i \sim \mathbb{P}(\mathbf{o}), \pi_i} \left[\left. \underset{t=0}{\overset{\infty}{\sum}} \gamma^{t} c_i(\mathbf{o}_{i, t}, \mathbf{a}_{i, t}) \right\rvert \mathbf{o}_{i,0} = \mathbf{o}_i \right],
\end{equation}
\hrulefill
\end{figure*}
\else
\fi
where $k_1$ and $k_2$ are system hyperparameters. The \ac{ppo} algorithm is summarized in Algorithm \ref{algo:ppo_algo}. We use two \acp{dnn}, one to approximate the policy $\pi(a\lvert s)$, which takes the state as an input and outputs a probability distribution over $\mathcal{A}$ and the action $a$ is sampled from this distribution. This network is trained to maximize the \ac{ppo} surrogate performance measure in (\ref{eq:ppo_surrogate_final}). The second \ac{dnn} approximates $V_{\pi}(s)$, and is trained to minimize the mean squared error between the output of the network and the average value of the state observed so far. The \ac{dnn} architecture used by all of the algorithms considered in this paper is described in detail on Appendix \ref{app:dnn_arch}.
\begin{algorithm}[!t]
\label{algo:ppo_algo}
\SetAlgoLined
\SetArgSty{}
\SetKw{Variables}{Variable Definition}
\SetKw{Init}{Initialization}
\SetKw{Out}{Output}
\SetKwRepeat{Loop}{loop}{end}
\caption{\ac{ppo} algorithm}
\SetKwProg{Fn}{\textbf{function}}{}{end}
\Init{
\begin{enumerate}
    \item Set learning rate $\alpha \in [0,1)$
    \item Set the update period $T$
    \item Set $\epsilon$, $k_1$, $k_2$
    \item Initialize $\mathbf{w}_{\text{old}}$ randomly
    \item Set $s = s_0 \in \mathcal{S}$
    \item Set $t \leftarrow 0$
\end{enumerate}
}
\Loop{ }{
$\mathcal{E} \leftarrow$ Initialize with an empty array of size $T$\;
\For{$m = 1 ... T$}{
    $a^\prime \sim \pi_{\mathbf{w}_{\text{old}}}(a|s)$\;
    $s^\prime \sim \mathbb{P}(s^\prime|s, a)$\;
    $c^\prime = c(s,a)$ \;
    $\mathcal{E}_m \leftarrow (s, a^\prime, c^\prime, s^\prime)$\;
}
\For{$n = 1 ... N_{\text{epochs}}$}{
    Sample minibatch $\tilde{\mathcal{E}}$ from $\mathcal{E}$ such that $|\tilde{\mathcal{E}}| < T$\;
    $\nabla_{\mathbf{w}} \tilde{J}(\mathbf{w}) \leftarrow \hat{E}_{\tilde{\mathcal{E}}} \left[ \nabla_{\mathbf{w}} \ln \pi_{\mathbf{w}}(a|s) A(s_t,a_t) \right]$ \;
    $\mathbf{w} \leftarrow \mathbf{w}_{\text{old}} + \alpha \nabla_{\mathbf{w}} \tilde{J}^{\text{surr}}(\mathbf{w})$\;
}
$\mathbf{w}_{\text{old}} \leftarrow \mathbf{w}$
}
\end{algorithm}

\subsection{Independent Learners}
\label{sec:full_dist}
In the \ac{il} architecture, each device has its own set of weights $\mathbf{w}_i$ and is running its own learning algorithm to update their weights without sharing information about their policies or current and previous states. As each device has only a local view of the state of the environment, it cannot learn the optimal joint-policy in (\ref{eq:posg_solution}). Therefore, each \ac{mtd} tries to find its local optimal policy as
\ifCLASSOPTIONtwocolumn
defined on (\ref{eq:il_posg}), on the top of the next page, 
\else
\begin{equation}
    \label{eq:il_posg}
    \pi^{*}_i = \arg \underset{\pi_i \in \Pi^{IL}_i}{\min} \text{ } E_{\mathbf{o}_i \sim \mathbb{P}(\mathbf{o}), \pi_i} \left[\left. \underset{t=0}{\overset{\infty}{\sum}} \gamma^{t} c_i(\mathbf{o}_{i, t}, \mathbf{a}_{i, t}) \right\rvert \mathbf{o}_{i,0} = \mathbf{o}_i \right],
\end{equation}
\fi
where $\Pi^{IL}_i = \{ \pi \lvert \pi: \mathcal{A}_i \times \mathcal{O}_i \rightarrow [0,1] \}$ is the set of all possible policies mapping the action-observation space into a probability. Notice that the joint policy search space of the \ac{il} \ac{posg} is $\bm{\Pi}^{IL} = \Pi^{IL}_1 \times \Pi^{IL}_2 \times \cdots \times \Pi^{IL}_{N_u}$. Additionally, the local cost function is given by
\ifCLASSOPTIONtwocolumn
\begin{eqnarray}
    \label{eq:il_cost}
    c_i(\mathbf{o}_{i}, \mathbf{a}_{i}) &=& x_i (P_{ON} + p_i) + (1 - x_i) P_{OFF} \nonumber \\ &&+ \omega_i \left( {b_i +  \mu_i \xi_i} \right).
\end{eqnarray}
\else
\begin{equation}
    \label{eq:il_cost}
    c_i(\mathbf{o}_{i}, \mathbf{a}_{i}) = x_i (P_{ON} + p_i) + (1 - x_i) P_{OFF} + \omega_i \left( {b_i +  \mu_i \xi_i} \right).
\end{equation}
\fi
Consequently, the local cost functions leads to the definition of a local value function
\begin{equation}
    \label{eq:il_value_func}
    V^{IL}_{\pi_i}(o_i) = E_{\pi_i} \left[\left. \underset{t = 0}{\overset{\infty}{\sum}} - \gamma^t c(o_{i,t}, a_{i,t}) \right\lvert o_{i, 0} = o_i\right].
\end{equation}
Furthermore, the policy function is approximated by a \ac{dnn} $\pi_{\mathbf{w}_i}$ that is trained on its previous experience using Algorithm \ref{algo:ppo_algo}. As both the policy and value \acp{dnn} are trained on the same \ac{mtd}, both the actor and the critic networks share the same weights to reduce the memory footprint, but have different output heads, the actor head outputs the probabilities of selecting each action, while the critic head outputs critic values. The diagram in Fig. \ref{fig:ann_diag} illustrates this architecture.

Effectively, each agent tries to solve the problem defined in (\ref{eq:il_posg}) while ignoring the effects of other agents, treating it as part of the environment. So, the problem reduces to a \ac{mdp} \cite{Puterman2014}. The agents change their policies independently of one another, but their actions affect the costs experienced by other agents. Therefore, the agents perceive the environment as non-stationary \cite{Omidshafiei2017a}. To the best of our knowledge, there are no known algorithms that give theoretical guarantees of convergence and optimality in the non-stationary \ac{mdp} setting nor on the solution of the general \ac{posg} problem posed in (\ref{eq:posg_solution}). However, the \ac{il} is considered to be a reasonable heuristic to find sub-optimal solutions to a \ac{posg} \cite{Matignon2012}.
\ifCLASSOPTIONtwocolumn
\begin{figure}
    \centering {
    \includegraphics[width=\columnwidth]{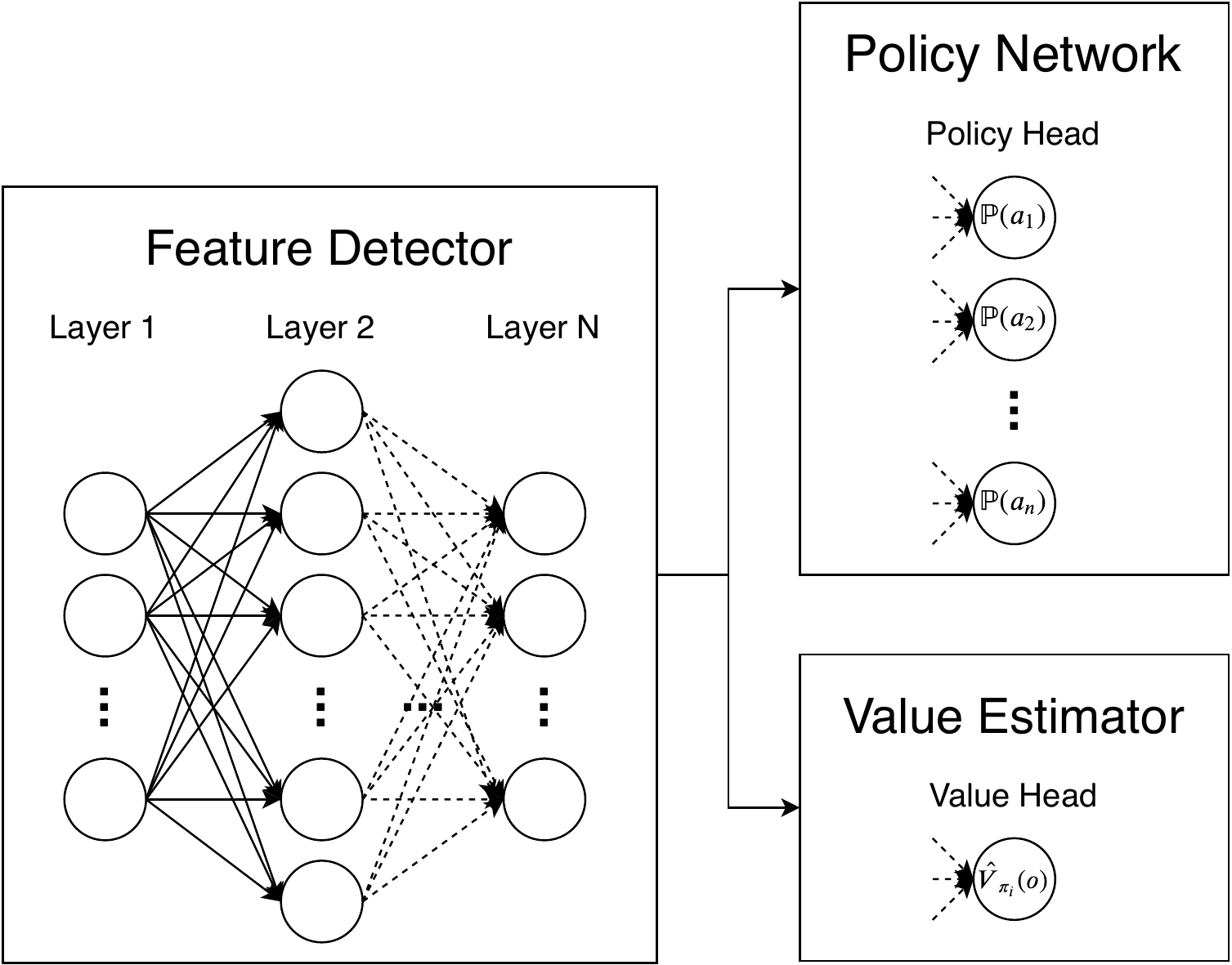}
    \caption{Diagram of \ac{dnn} architecture with shared weights and split actor and critic heads.}
    \label{fig:ann_diag}
    }
\end{figure}
\else
\begin{figure}
    \centering {
    \includegraphics[width=0.48\columnwidth]{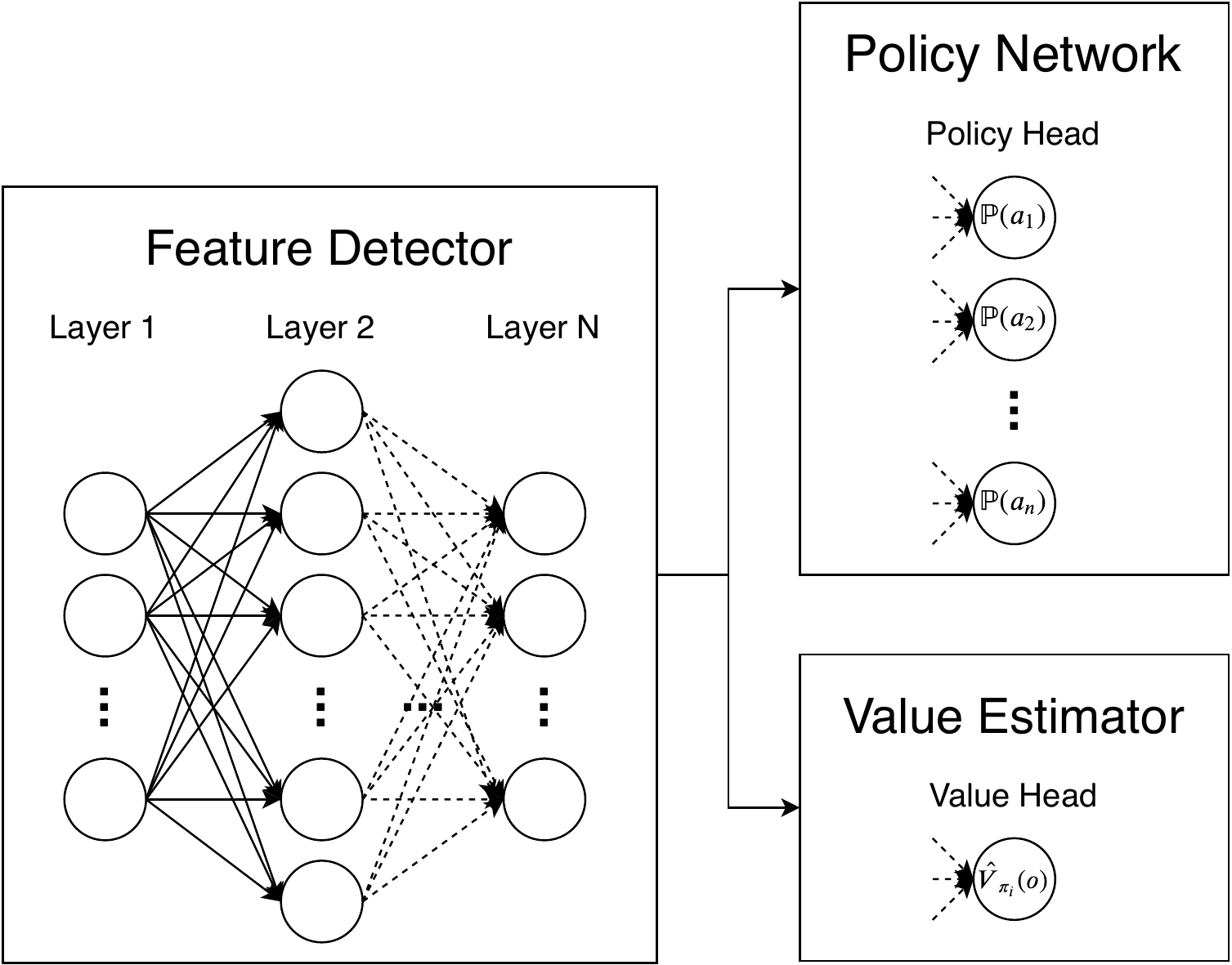}
    \vspace*{-.7cm}
    \caption{Diagram of \ac{dnn} architecture with shared weights and split actor and critic heads.}
    \label{fig:ann_diag}
    }
\end{figure}
\fi
As shown in Fig \ref{fig:arch_comp}, the main advantage of this approach is that it does not require any form of communications between devices nor between a device and the \ac{bs}. On the other hand, it requires every device to have its own set of weights and to run its own learning algorithm, which can result in a high power consumption. Also, as each agent faces a non-stationary environment, there are no guarantees of convergence to an optimal solution.

\subsection{Distributed Actor with Central Critic}
\label{sec:central_critic}
The \ac{ppo} algorithm makes use of two networks: the actor, which models the agent's policy, and the critic, which estimates the value of a state. Originally, the algorithm proposes that both networks can share weights to accelerate convergence and reduce memory costs \cite{Schulman2017}; In the \ac{dacc} architecture, each agent learns its own policy based on its local cost, similar to the \ac{il} architecture, while a single critic is stored and trained on an \revTwo{edge} computing node. The goal of this architecture is to mitigate the effects of the partial observation by having a critic that has access to the data of all the agents (the whole state) to estimate the value of the whole state $\mathbf{s}$, defined in (\ref{eq:value_func}), and not only the local value based on the local observations as done by the \ac{il} architecture. 
Thus, the \ac{dacc} \ac{posg} problem is given by
\ifCLASSOPTIONtwocolumn
(\ref{eq:dacc_posg}), located on top of next page, 
\begin{figure*}[ht!]
\begin{equation}
    \label{eq:dacc_posg}
    \pi^{*}_i = \arg \underset{\pi_i \in \Pi^{DACC}_i}{\min} \text{ } E_{\mathbf{o}_i \sim \mathbb{P}(\mathbf{o}), \pi_i} \left[\left. \underset{t=0}{\overset{\infty}{\sum}} \gamma^{t} c_i(\mathbf{o}_{i, t}, \mathbf{a}_{i, t}) \right\rvert \mathbf{o}_{i,0} = \mathbf{o}_i \right],
\end{equation}
\hrulefill
\end{figure*}
\else
\begin{equation}
    \label{eq:dacc_posg}
    \pi^{*}_i = \arg \underset{\pi_i \in \Pi^{DACC}_i}{\min} \text{ } E_{\mathbf{o}_i \sim \mathbb{P}(\mathbf{o}), \pi_i} \left[\left. \underset{t=0}{\overset{\infty}{\sum}} \gamma^{t} c_i(\mathbf{o}_{i, t}, \mathbf{a}_{i, t}) \right\rvert \mathbf{o}_{i,0} = \mathbf{o}_i \right],
\end{equation}
\fi
where $\Pi^{DACC}_i = \{ \pi \lvert \pi: \mathcal{A}_i \times \mathcal{O}_i \rightarrow [0,1] \}$ is the set of all possible probability distributions over the action-ovservation space, and the joint policy search space of the \ac{dacc} \ac{posg} is $\bm{\Pi}^{DACC} = \Pi^{DACC}_1 \times \Pi^{DACC}_2 \times \cdots \times \Pi^{DACC}_{N_u}$. While the critic value is computed on local observation data in the \ac{il} architecture, as shown in (\ref{eq:il_value_func}), the critic value in the \ac{dacc} architecture is computed over global state information, i.e. $V^{DACC}_{\bm{\pi}}(\mathbf{s}) = V_{\bm{\pi}}(\mathbf{s})$.

Both the policy function $\pi_i$ and the value function estimator $V^{DACC}_{\bm{\pi}}(\mathbf{s})$ are approximated by \acp{dnn}. The policy \ac{dnn} $\pi_{\mathbf{w}_i}$ is trained and stored on each device, while the value function estimator is stored and computed on an edge computing node. Hence, in this architecture, the surrogate objective function in (\ref{eq:ppo_surrogate_final}) is split into two, with one to be minimized by the devices to train the policy network, given by
\begin{equation}
\label{eq:ppo_surrogate_actor}
    J_a^{\text{surr}}(\mathbf{w}_i) = J^{\text{clip}}(\mathbf{w}_i) + k_2 H(\pi_{\mathbf{w}_i}),
\end{equation}
and the other to be minimized on the \revTwo{edge} to train the value function network, given by
\begin{equation}
\label{eq:ppo_surrogate_critic}
    J_c^{\text{surr}}(\mathbf{w_c}) = J^{\text{VF}}(\mathbf{w_c}).
\end{equation}
%
Furthermore, as illustrated in Fig. \ref{fig:arch_comp}, each agent keeps its own set of weights $\mathbf{w}_i$ for the actor network, while the weights of the value function estimator $\mathbf{w}_{c}$ are stored and updated on the edge computing node. Additionally, both the \acp{mtd} and the edge node must perform backpropagation to update their weights. While each \ac{mtd} has access to its own local information, the value estimator trained on the edge can leverage the data collected by all agents, and thus, the edge agent is able to backpropagate on the global state information.

Moreover, as shown in (\ref{eq:perf_gradient}), the critic value is necessary to compute the \ac{ppo} gradient. Therefore, this architecture requires the \acp{bs} to feedback the value of each state, $V_{\bm{\pi}}(\mathbf{s})$ given in (\ref{eq:value_func}), after every \ac{tti}, such that the agents are able to perform backpropagation and train their policy networks. Moreover, while the channel response can be estimated by the network from the preambles, the buffer occupancy information $b_i$ needs to be sent by the \acp{mtd} to the edge in every \ac{tti}, thus, the edge node is able to compute the value functions and its weight's update.

\subsection{Centralized Learning with Distributed Inference}
\label{sec:central_learn}
As the number of \acp{mtd} in the network increases, the size of the policy search space for the \ac{il} and \ac{dacc} architectures increase exponentially, consequently increasing the solution space. To address this issue, in the \ac{cldi} architecture, there is a single set of weights, and therefore a single policy $\pi$ and a search space $\Pi^{CLDI} = \{ \pi \lvert \pi: \mathcal{A}_i \times \mathcal{O}_i \rightarrow [0,1], \text{ } i = 1, \dots N_U \}$ that does not increase in size with the number of \acp{mtd}. Both the policy and the critic are trained on the \revTwo{edge} and an updated set of weights is periodically broadcast to the \acp{mtd}, thus reducing the computational burden required to train a neural network on the devices. Moreover, the policy on the \revTwo{edge} is trained on data from all \acp{mtd} leading to improved sample efficiency. Hence, instead of solving (\ref{eq:posg_solution}), the \ac{cldi} architecture looks for solutions to
\ifCLASSOPTIONtwocolumn
(\ref{eq:cldi_posg}), defined on the top of the following page, 
\begin{figure*}[!ht]
\begin{equation}
    \label{eq:cldi_posg}
    \pi^{*} = \arg \underset{\pi \in \Pi^{CLDI}}{\min} \text{ } E_{\mathbf{s} \sim \mathbb{P}(\mathbf{s}), \pi} \left[\left. \underset{t=0}{\overset{\infty}{\sum}} \gamma^{t} c_{CLDI}(\mathbf{s}_{t}, \mathbf{a}_{t}) \right\rvert \mathbf{s}_{0} = \mathbf{s} \right],
\end{equation}
\hrulefill
\end{figure*}
\else
\begin{equation}
    \label{eq:cldi_posg}
    \pi^{*} = \arg \underset{\pi \in \Pi^{CLDI}}{\min} \text{ } E_{\mathbf{s} \sim \mathbb{P}(\mathbf{s}), \pi} \left[\left. \underset{t=0}{\overset{\infty}{\sum}} \gamma^{t} c_{CLDI}(\mathbf{s}_{t}, \mathbf{a}_{t}) \right\rvert \mathbf{s}_{0} = \mathbf{s} \right],
\end{equation}
\fi
where the \ac{cldi} cost function is given by
\ifCLASSOPTIONtwocolumn
\begin{eqnarray}
    \label{eq:cldi_cost}
    c_{CLDI}(\mathbf{s}, \mathbf{a}) = \frac{1}{N_U} \underset{i = 1}{\overset{N_U}{\sum}}& x_i (P_{ON} + p_i) \nonumber \\ &+ (1 - x_i) P_{OFF}  \nonumber \\ 
    &+ \omega_i \left( {b_i +  \mu_i \xi_i} \right),
\end{eqnarray}
\else
\begin{equation}
    \label{eq:cldi_cost}
    c_{CLDI}(\mathbf{s}, \mathbf{a}) = \frac{1}{N_U} \underset{i = 1}{\overset{N_U}{\sum}} x_i (P_{ON} + p_i) + (1 - x_i) P_{OFF} + \omega_i \left( {b_i +  \mu_i \xi_i} \right),
\end{equation}
\fi
which is the average cost function of the \acp{mtd}. It is worth highlighting that the \ac{cldi} cost is an average of the costs of all \acp{mtd}, thus, the shared policy is updated to increase the average performance of all \acp{mtd}, as opposed to the \ac{il} and the \ac{dacc} architecture where the policy of each \ac{mtd} is updated to optimize its local performance. Both the policy and value function networks are stored and trained on the edge following Algorithm \ref{algo:ppo_algo} using the cost function defined in (\ref{eq:cldi_cost}). The devices have a copy of the policy network, but they do not train it, they just use it for decision-making. In this architecture, the devices must append the buffer state information to every transmitted packet, and thus, the networks can be trained on the edge node, where the network must send the updated weights back to the \acp{mtd} periodically.

\section{Numerical Experiments}
\label{sec:numerical_exps}
In this section, the performance of the proposed architectures is evaluated through computer simulations. In order to provide a frame of reference, we also simulate the performance of a baseline employing a reactive \ac{harq} protocol with power boosting. The details of the baseline are described in Appendix \ref{sec:baseline}. We consider that there are two \acp{bs} and eight subcarriers serving a circular area with a $300$ m radius. We generate $1000$ realizations of this scenario, where at each realization we place both the \acp{bs} and the \acp{mtd} in a random location within the circular area. At each realization the learning algorithms start from scratch (e.g. the weights of the agents are randomly initialized at the beginning of each realization) and runs for $15000$ \ac{tti}. Then, we compare the average performances, along with their variances, with respect to the average delay experienced by the network, the number of dropped packets, the average power spent, and the number of collisions. 
\begin{table}[!t]
\begin{center}
\caption{Parameters used in the simulations \label{tab:parameters}}
\vspace{-.5cm}
\begin{tabular}{ c | c | c | c}
\hline
Parameter & Value & Parameter & Value \\
\hline
\hline
$f_S$ & $10^5$ symbols/s & $L$ & $100$ bytes\\
$R$   & $300$ m & $\Delta_t$ & $10$ ms\\
$N_U$ & $\{2560, 7680\}$ users & $\delta_i$ & $U \left(\{ 4, 8, 12 \}\right)$ packets\\
$N_B$ & $2$ \ac{bs} & $\lambda_i$ & $U \left( \{40, 60, 80\} \right)$ packets/s\\
$N_S$ & $8$ subcarriers & $\gamma$ & $0.99$\\
$N_P$ & $64$ preambles & $P_{ON}$ & $320$ milliwatts\\
$\alpha$ & $3.5$ & $P_{OFF}$ & $0$ milliwatts\\
$B$ & $25$ packets & $f_{\max}$ & $10$ Hz\\
T & $200$ \ac{tti} & & \\
\end{tabular}
\end{center}
\end{table}
\subsection{Results}
\label{sec:results}
\ifCLASSOPTIONtwocolumn
\begin{figure*}
    \centering {
    \includegraphics[width=1.7\columnwidth]{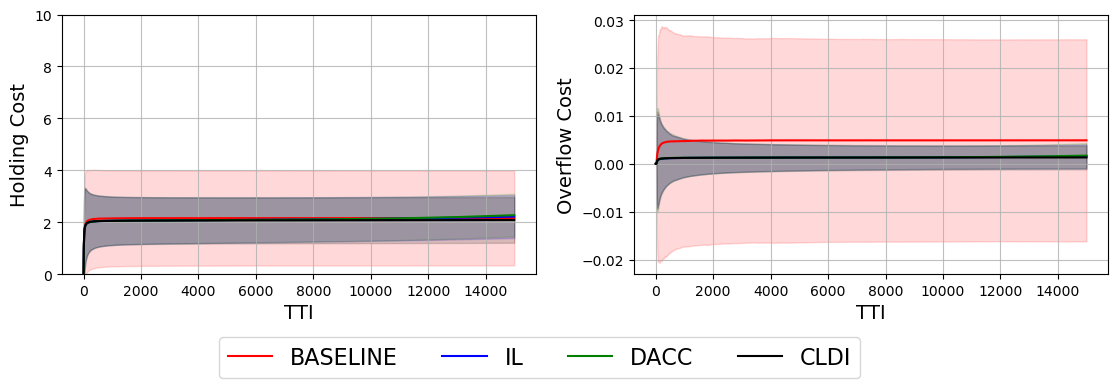}
    \caption{Simulation results showing the holding costs and overflow costs with $2560$ \acp{mtd} in the simulated area for the three proposed architectures and the baseline, where \ac{il} stands for the independent learners, \ac{dacc} for distributed actor with central critic and \ac{cldi} for central learning with decentralized inference.}
    \label{fig:results_Nu40a}
    }
\end{figure*}
\else
\begin{figure}
    \centering {
    \includegraphics[width=.85\columnwidth]{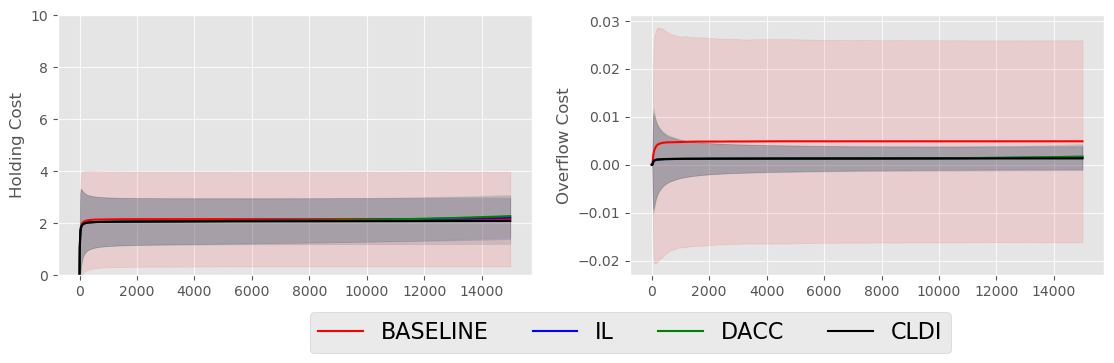}
    \vspace*{-.75cm}
    \caption{Simulation results showing the holding costs and overflow costs with $2560$ \acp{mtd} in the simulated area for the three proposed architectures and the baseline, where \ac{il} stands for the independent learners, \ac{dacc} for distributed actor with central critic and \ac{cldi} for central learning with decentralized inference.}
    \label{fig:results_Nu40a}
    }
\end{figure}
\fi
\ifCLASSOPTIONtwocolumn
\begin{figure*}
    \centering {
    \includegraphics[width=1.7\columnwidth]{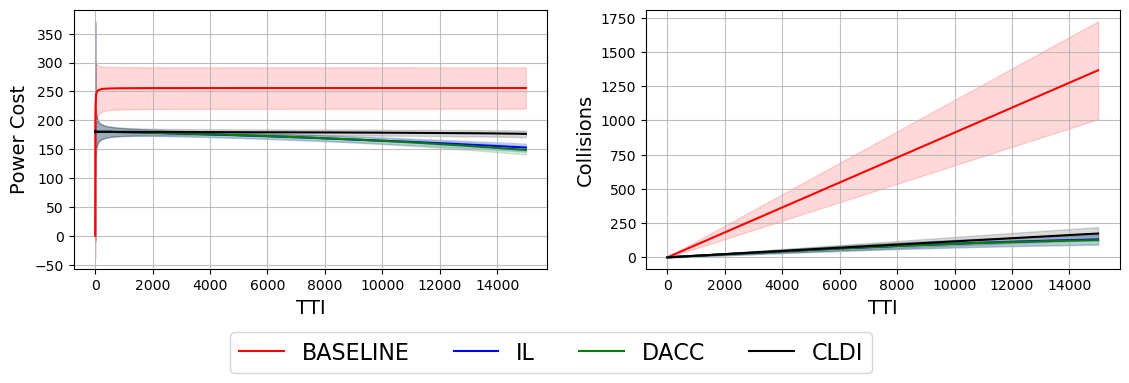}
    \caption{Simulation results showing the power costs and the number of collisions with $2560$ \acp{mtd} in the simulated area for the three proposed architectures and the baseline, where \ac{il} stands for the independent learners, \ac{dacc} for distributed actor with central critic and \ac{cldi} for central learning with decentralized inference.}
    \label{fig:results_Nu40b}
    }
\end{figure*}
\else
\begin{figure}
    \centering {
    \includegraphics[width=.85\columnwidth]{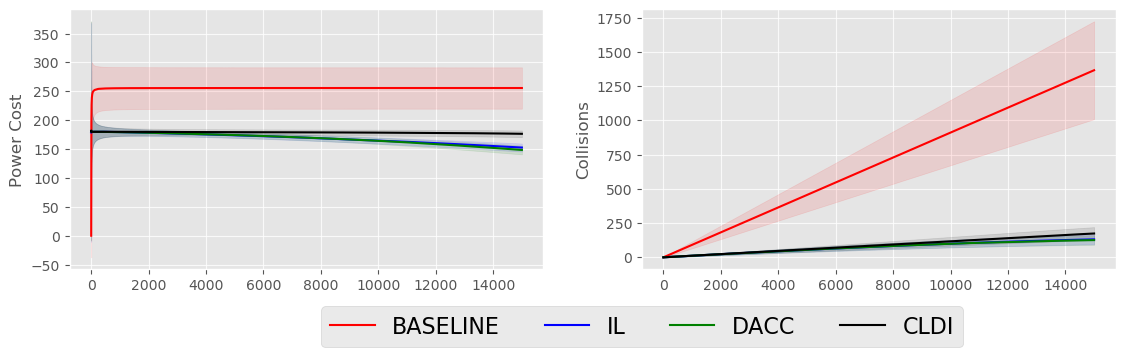}
    \vspace*{-.5cm}
    \caption{Simulation results showing the power costs and the number of collisions with $2560$ \acp{mtd} in the simulated area for the three proposed architectures and the baseline, where \ac{il} stands for the independent learners, \ac{dacc} for distributed actor with central critic and \ac{cldi} for central learning with decentralized inference.}
    \label{fig:results_Nu40b}
    }
\end{figure}
\fi
We compare the baseline and the architectures proposed in Section \ref{sec:learning_arch} in terms of the average network delay, power, dropped packets, and collisions during $15000$ \acp{tti}. We evaluate the network delay through the holding cost, as the average network delay is proportional to the number of packets held in the devices' buffer. We consider that devices with different mean packet arrival rates and latency constraints are being serviced by the same cellular network. For each realization, the packet arrival rate of each \acp{mtd} is uniformly sampled from $\{40, 60, 80\}$ packets per second and the latency constraint is uniformly sampled from $\{ 4, 8, 12\}$ queued packets. \revOne{Notice that in all the plots the $x$-axis shows the \ac{tti}. In the holding cost plot, the $y$-axis shows the cumulative average of the number of packets in the buffer at a given \ac{tti}. The $y$-axis in the overflow cost plots show the cumulative average value of $\zeta_i$. Furthermore, the $y$-axis in the power cost plots shows the cumulative average of the power spent by \acp{mtd} in milliwatts. Finally, the $y$-axis in the collisions' plot shows the cumulative sum of collisions up to the given \ac{tti}.}

As shown in Fig. \ref{fig:results_Nu40a}, with $2560$ users, the average holding cost between all four approaches is roughly the same. However, we notice that the baseline presents a significantly higher variance than the proposed architectures. Furthermore, the average network delay is below four, which is the smallest latency constraint in the network, within at least one standard deviation. With respect to overflown packets, also in Fig. \ref{fig:results_Nu40a}, on average the baseline approach drops slightly more packets than the proposed architectures, but again with significantly more variance. With respect to the power consumption, as shown in Fig. \ref{fig:results_Nu40b}, the three proposed architectures spend on average roughly $70\%$ of the power spent by the baseline. Moreover, as mentioned in Section \ref{sec:learning_arch}, the \ac{cldi} algorithm tends to converge faster as it is trained on observations from every device in the network and has to search for a policy in a notably small policy search space. This is confirmed by the fact that, as the simulation advances in time and the \ac{il} and \ac{dacc} algorithms train on more data, they achieve similar performance levels as \ac{cldi}, while using less power. The performance improvement of the proposed architectures compared to the baseline is even more significant when it comes to the number of collisions, as shown in Fig. \ref{fig:results_Nu40b}. On average, the reinforcement learning based solutions experience $15\%$ of the baseline's collisions during the same period of time.

\revTwo{Moreover, in all investigated architectures, the holding cost performance, when averaged over users with the same delay constraint, follows the same trend as when averaged over all users (shown in Fig. \ref{fig:results_Nu40a}). Therefore, we can conclude that in a scenario with $2560$ \acp{mtd}, in average, all the architectures satisfy the delay constraints. However, the baseline approach presents larger performance fluctuations, as shown by the larger standard deviation in Fig. \ref{fig:results_Nu40a}.} 
\ifCLASSOPTIONtwocolumn
\begin{figure*}
    \centering {
    \includegraphics[width=1.7\columnwidth]{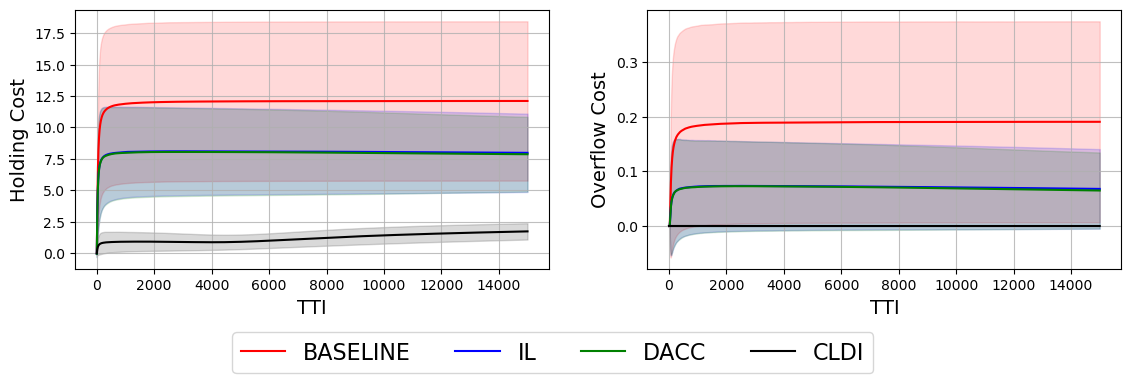}
    \caption{Simulation results showing the holding costs and overflow costs with $7680$ \acp{mtd} in the simulated area for the three proposed architectures and the baseline, where \ac{il} stands for the independent learners, \ac{dacc} for distributed actor with central critic and \ac{cldi} for central learning with decentralized inference.}
    \label{fig:results_Nu120a}
    }
\end{figure*}
\else
\begin{figure}
    \centering {
    \includegraphics[width=.85\columnwidth]{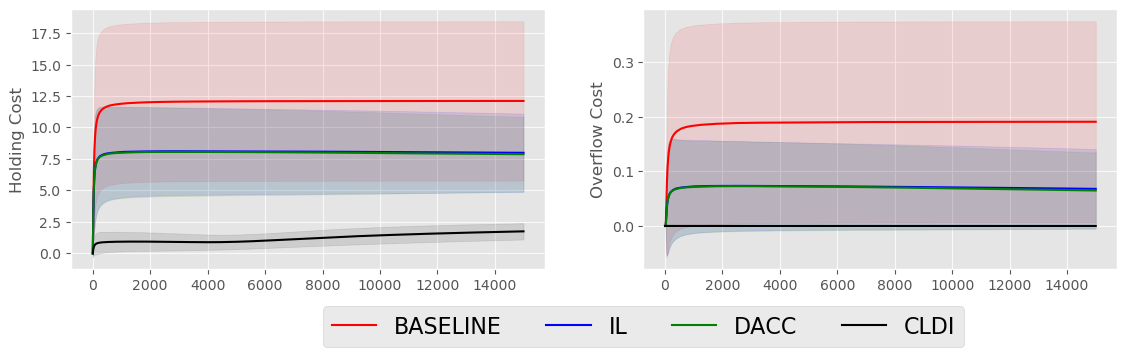}
    \vspace*{-.5cm}
    \caption{Simulation results showing the holding costs and overflow costs with $7680$ \acp{mtd} in the simulated area for the three proposed architectures and the baseline, where \ac{il} stands for the independent learners, \ac{dacc} for distributed actor with central critic and \ac{cldi} for central learning with decentralized inference.}
    \label{fig:results_Nu120a}
    }
\end{figure}
\fi
\ifCLASSOPTIONtwocolumn
\begin{figure*}
    \centering {
    \includegraphics[width=1.7\columnwidth]{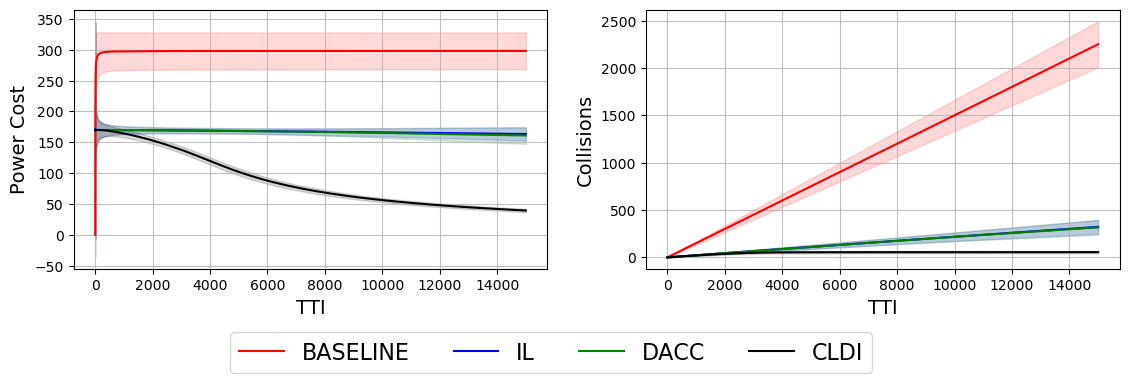}
    \caption{Simulation results showing the power costs and the number of collisions with $7680$ \acp{mtd} in the simulated area for the three proposed architectures and the baseline, where \ac{il} stands for the independent learners, \ac{dacc} for distributed actor with central critic and \ac{cldi} for central learning with decentralized inference.}
    \label{fig:results_Nu120b}
    }
\end{figure*}
\else
\begin{figure}
    \centering {
    \includegraphics[width=.85\columnwidth]{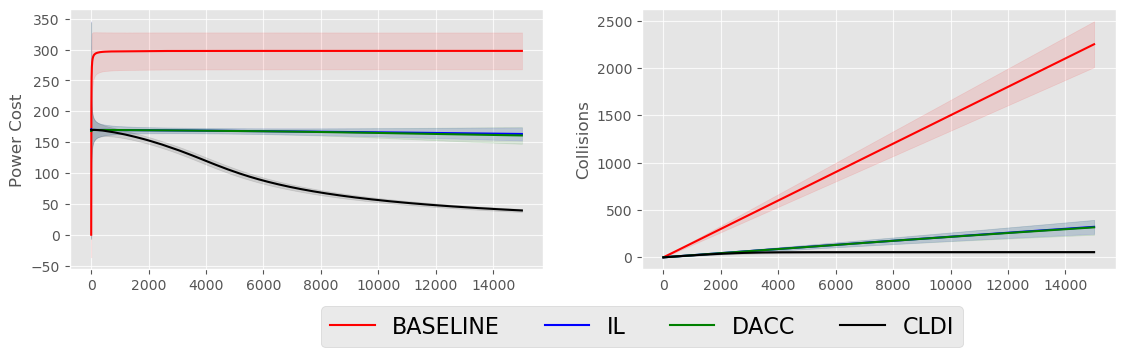}
    \vspace*{-.5cm}
    \caption{Simulation results showing the power costs and the number of collisions with $7680$ \acp{mtd} in the simulated area for the three proposed architectures and the baseline, where \ac{il} stands for the independent learners, \ac{dacc} for distributed actor with central critic and \ac{cldi} for central learning with decentralized inference.}
    \label{fig:results_Nu120b}
    }
\end{figure}
\fi

As illustrated in Fig. \ref{fig:results_Nu120a}, when the number of users is increased to $7680$, the average holding cost of the \ac{cldi} architecture converges to $2$ packets, while the \ac{il} and \ac{dacc} converge to $8$ packets and the baseline to $12$ packets. From this result, we conclude that as the number of users increases the lack of collaboration between the \acp{mtd} in the \ac{il} and \ac{dacc} architectures starts to impact the average network delay, while \ac{cldi} performance stays around the same as for $2560$ users. Also in Fig. \ref{fig:results_Nu120a}, the average overflow cost of \ac{cldi} still remains around $0$, while the \ac{il} and \ac{dacc} estabilize around $0.7$ and the baseline at $0.19$. With regards to the average power costs at convergence, the \ac{cldi} architecture spends $16.66\%$ of the power spent by the baseline, while the \ac{il} and \ac{dacc} spend $52\%$, as seen in Fig. \ref{fig:results_Nu120b}. The significant decrease in the power spent by \ac{cldi} is explained by the centralized training, which makes more training data available, since \ac{cldi} has $7680$ new data points for each \ac{tti} while the other architectures have only $1$, which points to a cooperative behavior arising among the \acp{mtd}. This is also reflected in the collisions performance, where \ac{cldi} experiences around $2.25\%$ of the baseline's collisions and \ac{il} and \ac{dacc} experience around $14\%$.

\revTwo{Furthermore, similar to the $2560$ \acp{mtd} case,  in all architectures investigated, all devices converge to roughly the same average holding cost, regardless of the delay constraint. Thus, in the $7680$ \acp{mtd} scenario, only the \ac{cldi} architecture maintains an average holding cost below the delay constraints for devices with $\delta_i = 4$, $\delta_i = 8$ and $\delta_i = 12$. In the \ac{il} and \ac{dacc} architectures, in average, only devices with $\delta_i = 8$ and $\delta_i = 12$ satisfy their constraints. Finally, when the baseline architecture is employed, on average, none of the \acp{mtd} is able to satisfy its constraint. This confirms that the \ac{cldi} architecture scales better than the others in densely deployed scenarios.} 
\subsection{Tradeoffs}
\label{sec:tradeoff}

\ifCLASSOPTIONtwocolumn
\begin{table*}[!t]
\begin{center}
\caption{Overhead and Performance Tradeoffs for $N_U = 2560$ \label{tab:tradeoff_Nu40}}
\begin{tabular}{ c | c  c  c  c  c  c }
\hline
Algorithm & DL Overhead & UL Overhead & Collisions & Power Cost & Holding Cost \\
\hline
\hline
Baseline & - & - & $1364$ collisions & $255.54$ mW & $2.15$ packets  \\
\ac{il} & - & - & $131$ collisions & $168.68$ mW & $2.11$ packets  \\
\ac{dacc} & $1.6$ kbits/s & $1.6$ kbits/s & $122$ collisions & $168.06$ mW & $2.12$ packets  \\
\ac{cldi} & $20.496$ kbits/s  & $1.6$ kbits/s & $174$ collisions & $178.81$ mW & $2.07$ packets \\
\end{tabular}
\end{center}
\end{table*}
\begin{table*}[!t]
\begin{center}
\caption{Overhead and Performance Tradeoffs for $N_U = 7680$ \label{tab:tradeoff_Nu120}}
\begin{tabular}{ c | c  c  c  c  c  c }
\hline
Algorithm & DL Overhead & UL Overhead & Collisions & Power Cost & Holding Cost \\
\hline
\hline
Baseline & - & - & $2253$ collisions & $297.6154$ mW & $12.01$ packets  \\
\ac{il} & - & - & $322$ collisions & $166.98$ mW & $8.025$ packets  \\
\ac{dacc} & $1.6$ kbits/s & $1.6$ kbits/s & $317$ collisions & $166.25$ mW & $7.95$ packets  \\
\ac{cldi} & $20.496$ kbits/s  & $1.6$ kbits/s & $55$ collisions & $87.81$ mW & $1.21$ packets \\
\end{tabular}
\end{center}
\end{table*}
\else
\begin{table}[!t]
\begin{center}
\caption{Overhead and Performance Tradeoffs for $N_U = 2560$
\vspace{-.5cm}
\label{tab:tradeoff_Nu40}}
\begin{tabular}{ c | c  c  c  c  c  c }
\hline
Algorithm & DL Overhead & UL Overhead & Collisions & Power Cost & Holding Cost \\
\hline
\hline
Baseline & - & - & $1364$ collisions & $255.54$ mW & $2.15$ packets  \\
\ac{il} & - & - & $131$ collisions & $168.68$ mW & $2.11$ packets  \\
\ac{dacc} & $1.6$ kbits/s & $1.6$ kbits/s & $122$ collisions & $168.06$ mW & $2.12$ packets  \\
\ac{cldi} & $20.496$ kbits/s  & $1.6$ kbits/s & $174$ collisions & $178.81$ mW & $2.07$ packets \\
\end{tabular}
\end{center}
\end{table}

\begin{table}[!t]
\begin{center}
\caption{Overhead and Performance Tradeoffs for $N_U = 7680$
\vspace{-.5cm}
\label{tab:tradeoff_Nu120}}
\begin{tabular}{ c | c  c  c  c  c  c }
\hline
Algorithm & DL Overhead & UL Overhead & Collisions & Power Cost & Holding Cost \\
\hline
\hline
Baseline & - & - & $2253$ collisions & $297.6154$ mW & $12.01$ packets  \\
\ac{il} & - & - & $322$ collisions & $166.98$ mW & $8.025$ packets  \\
\ac{dacc} & $1.6$ kbits/s & $1.6$ kbits/s & $317$ collisions & $166.25$ mW & $7.95$ packets  \\
\ac{cldi} & $20.496$ kbits/s  & $1.6$ kbits/s & $55$ collisions & $87.81$ mW & $1.21$ packets \\
\end{tabular}
\end{center}
\end{table}
\fi

In this subsection, we analyze the advantages and disadvantages of each of the proposed architectures, and discuss possible application scenarios.

The \ac{il} architecture does not require a central \revTwo{edge} entity to work, and therefore it cuts all the necessary overhead associated to data transmission between \acp{mtd} and \acp{bs}. However, each \ac{mtd} has to perform training and inference of its \ac{dnn}, which can be computationally expensive. Moreover, as each \ac{mtd} is trained in a fully distributed manner, without sharing any information, there is no chance of cooperation arising. Both \ac{dacc} and \ac{cldi} architectures require \acp{mtd} to transmit information about their number of packets currently in the buffers. On the other hand, since part of the training for \ac{dacc}, and all the training for \ac{cldi} is performed in the \revTwo{edge}, some of the computational burden is offloaded, thus saving power and easing the computation requirements of \acp{mtd}. In this work, we use $16$ bit floating point numbers to encode the state information, action, and network weights. Furthermore, we consider any extra data exchange that is not the payload to be overhead. Thus, in the uplink direction, the overhead of the \ac{dacc} and \ac{cldi} architectures is given by $\frac{16}{\Delta_t}$. Regarding the downlink overhead, in the \ac{dacc} architecture the network must send the critic value in every \ac{tti}, and therefore, the overhead is also $\frac{16}{\Delta_t}$. Meanwhile, in the \ac{cldi} architecture, the network must send all the weights of the \ac{dnn} to the \acp{mtd} every $200$ \ac{tti}.

Tables \ref{tab:tradeoff_Nu40} and \ref{tab:tradeoff_Nu120} show the average performance of each architecture and its respective overhead. As shown in Section \ref{sec:results}, for a smaller user density, the \ac{il} and \ac{dacc} architectures slightly outperform \ac{cldi}. However, for a higher density of \acp{mtd}, the \ac{cldi} architecture is able to leverage data from observations collected from all \acp{mtd}, and thanks to the centralized training, the \acp{mtd} work together to use the network resources more equitably, resulting in tremendous power savings, small average delays, and a minimal number of dropped packets and collisions. Therefore, we conclude that for cellular networks designed to serve a smaller number of \acp{mtd}, the \ac{il} and \ac{dacc} architectures, depending on whether the the devices have enough computational power to train their \ac{dnn} and how much overhead is tolerated, are recommended. However, for cellular networks designed to support a massive number of low-cost devices, the \ac{cldi} architecture is deemed more suitable.

\ifCLASSOPTIONtwocolumn
\begin{figure*}
    \centering {
    \includegraphics[width=2\columnwidth]{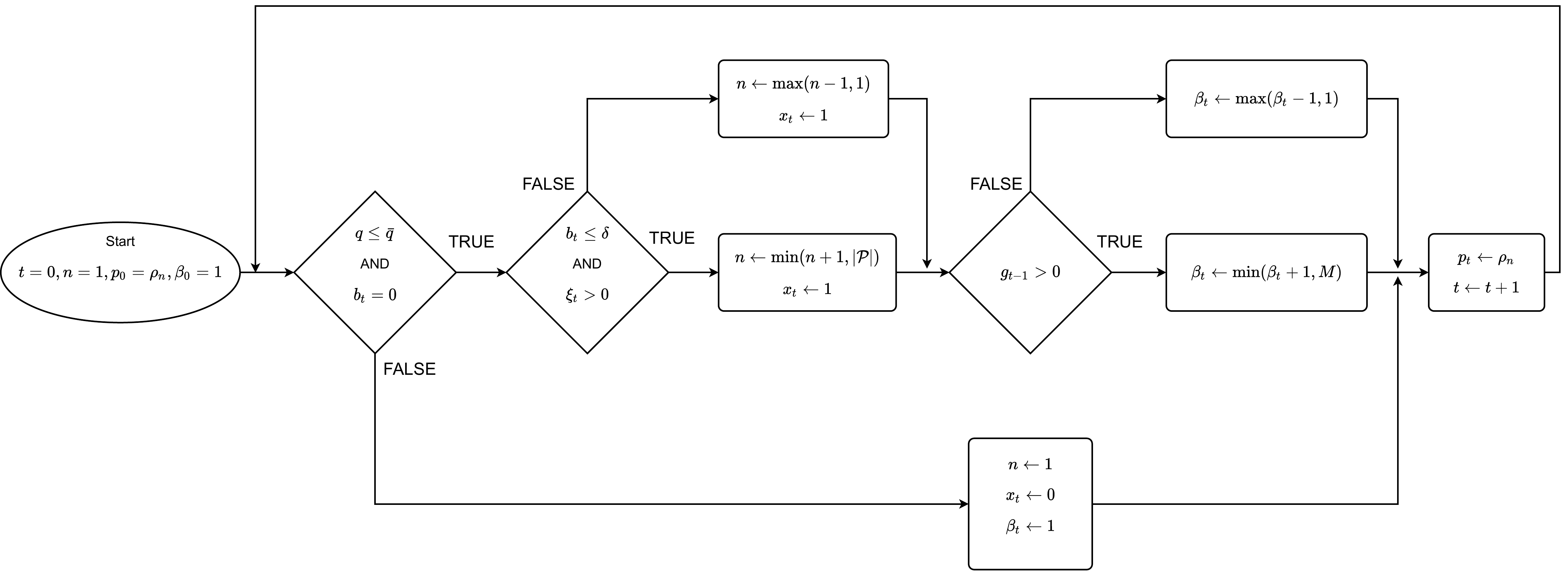}
    \caption{Flowchart of the baseline algorithm.}
    \label{fig:baseline}
    }
\end{figure*}
\else
\fi

\section{Conclusions}
\label{sec:conclusions}
\ifCLASSOPTIONtwocolumn
\begin{table*}[!t]
\begin{center}
\caption{Number of Weights in the \ac{dnn} \label{tab:weights}}
\begin{tabular}{ c  c  c  c}
\hline
\ac{gru} & Fully Connected Layers & Policy Network Head & Value Function Network Head \\
\hline
\hline
$3 [32^2 + 32 (4 + N_B N_S) + 32]$ & $2 \times 32^2$ & $32 2 M |\mathcal{P}| N_S$ & $32$ \\
\end{tabular}
\end{center}
\end{table*}
\fi

In this paper, we proposed a system model for \ac{mmtc} networks using grant-free transmission and formulated it as an average power minimization problem subject to delay constraints. Based on the related literature, we conclude that static access protocols are inefficient to handle the optimization problem and proposed three reinforcement learning based architectures to solve the optimization problem in a distributed fashion. The architectures have different degrees of centralization and overhead . Furthermore, we simulated the three architectures and compared their performance among against a static access policy baseline based on the reactive \ac{harq} protocol with power boosting. Finally, we showed that all three learnable architectures outperform the static baseline and we proceeded to analyze the tradeoffs between the architectures.

\appendices

\section{Baseline Algorithm}
\label{sec:baseline}
\ifCLASSOPTIONtwocolumn
\else
\begin{figure}
    \centering {
    \includegraphics[width=\columnwidth]{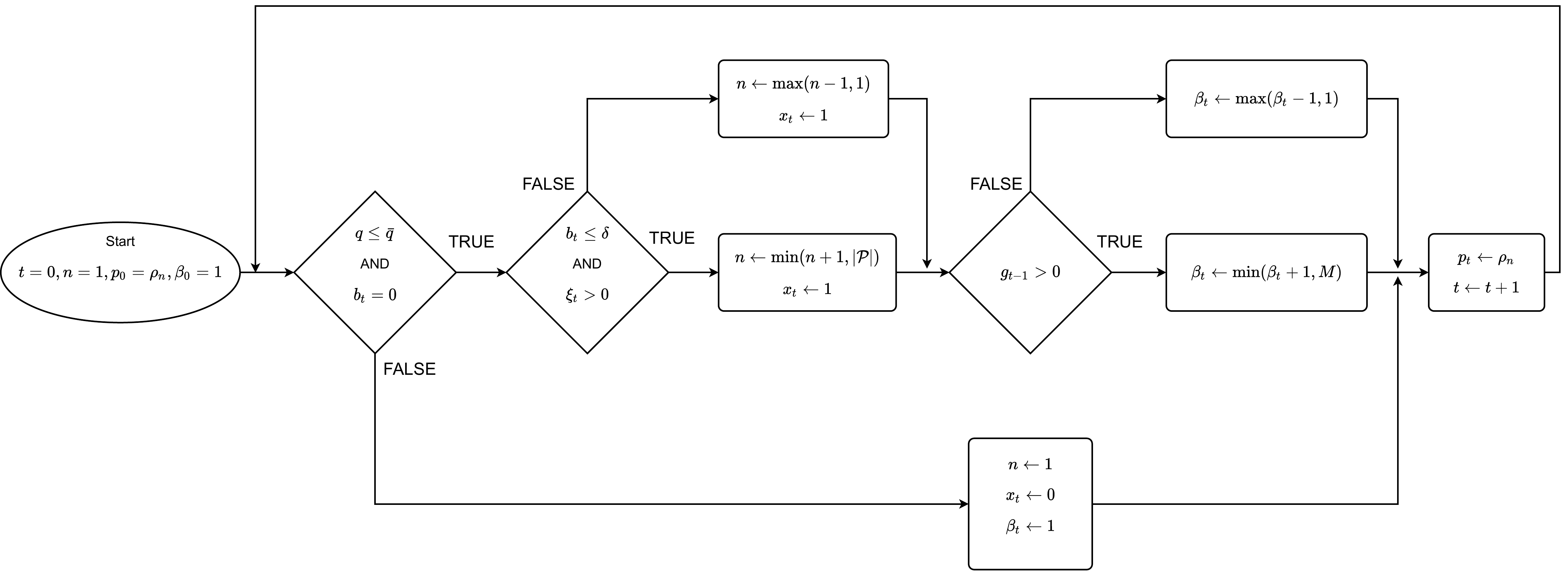}
    \vspace*{-.5cm}
    \caption{Flowchart of the baseline algorithm.}
    \label{fig:baseline}
    }
\end{figure}
\fi
Firstly, each device with packets in the buffer to transmit draws a random number $q \sim U \left( [0,1] \right)$, and if $q \leq \bar{q}$, where $\bar{q}$ is a congestion control threshold, the device tries to access the channel. This is done to avoid congestion by having all devices trying to access the channel at the same time. Furthermore, if the \ac{mtd} is currently violating its delay constraints or if there was a dropped packet in the last \ac{tti}, the device ramps up its power. Moreover, if at least one packet was successfully transmitted on the last \ac{tti}, the \ac{mtd} assumes it is facing a good channel condition, and it then increases the transmission modulation order. Otherwise, it assumes a bad channel and decreases it. The algorithm is described by the flowchart in Fig. \ref{fig:baseline}.


\section{\ac{dnn} Architecture and Parameters}
\label{app:dnn_arch}
\ifCLASSOPTIONtwocolumn
\else
\begin{table}[!t]
\begin{center}
\caption{Number of Weights in the \ac{dnn} \label{tab:weights}}
\vspace{-.5cm}
\begin{tabular}{ c  c  c  c}
\hline
\ac{gru} & Fully Connected Layers & Policy Network Head & Value Function Network Head \\
\hline
\hline
$3 [32^2 + 32 (4 + N_B N_S) + 32]$ & $2 (32^2 + 32)$ & $66 M |\mathcal{P}| N_S$ & $64$ \\
\end{tabular}
\end{center}
\end{table}
\fi

One of the requirements is that the \ac{dnn} must be shallow and relatively small to keep a light memory footprint on the devices and to reduce the computational complexity of the training and inference. We consider a \ac{gru} \cite{Cho2014} connected to a two-layer perceptron. As the observations of the \acp{mtd} are temporally correlated (through the number of packets in the buffer, and the channel gains) we include a \ac{gru} in the input to extract information from sequences of states. We employ \acp{gru} as it has been shown that they have comparable performance to the more commonly used \ac{lstm} units while being more computationally efficient \cite{Chung2014}. In our model, we consider a \ac{gru} with $N_B N_S + 4$ inputs, where $N_B N_S$ inputs take the channel state information, and the remaining four are the number of packets in the buffer ($b_i$), the number of arriving packets ($l_i$), the goodput on the previous \ac{tti} ($g_i$) and the number of overflown packets in the previous \ac{tti} ($\xi_i$). The \ac{gru} unit has $32$ output values, while both of the linear layers have $32$ inputs and $32$ outputs. Finally, the actor head has $32$ inputs and $2 M |\mathcal{P}| N_S$ outputs (one for each possible action), while the critic head has $32$ inputs and one output (the critic value). Table \ref{tab:weights} summarize the number of weights needed for each stage of the network\footnote{We used the values in \cite{Dey2017} to compute the number of weights needed by a \ac{gru}.}.

The networks are trained using an \ac{adam} optimizer \cite{Kingma2014} with a learning rate of $7 \times 10^{-4}$. At each \ac{dnn} network update, the weights are trained over $4$ \ac{ppo} epochs with $10$ minibatches per epoch. To avoid large gradient updates that make the optimization unstable, the gradients are clipped such that $\left \lVert \nabla J_{\mathbf{w}} \right \rVert \leq 0.5$. A value loss coefficient $k_1 = 0.5$ and an entropy loss coefficient $k_2 = 0.01$ are used.

\ifCLASSOPTIONcaptionsoff
  \newpage
\fi

\bibliographystyle{IEEEtran}
\bibliography{IEEEabrv.bib,references.bib}{}









\end{document}